\documentclass[%
 reprint,
superscriptaddress,
 amsmath,amssymb,
 aps,
prb,
]{revtex4-2}

\usepackage{amsmath}
\usepackage{amsfonts}
\usepackage{graphicx}
\usepackage{xcolor}
\usepackage{braket}
\usepackage[normalem]{ulem}
\usepackage{comment}
\usepackage{acro}

\usepackage{dcolumn}
\usepackage{bm}

\DeclareAcronym{MLIP}{short=MLIP, long=Machine Learning Interatomic Potential}
\DeclareAcronym{NLP}{short=NLP, long=Natural Language Processing}
\DeclareAcronym{MD}{short=MD, long=Molecular Dynamics}
\DeclareAcronym{DFT}{short=DFT, long=Density Functional Theory}
\DeclareAcronym{MEP}{short=MEP, long=minimum energy path}
\DeclareAcronym{CINEB}{short=CI-NEB, long=climbing image nudged elastic band}
\DeclareAcronym{NEB}{short=NEB, long=nudged elastic band}
\DeclareAcronym{ACE}{short=ACE, long=Atomic Cluster Expansion}
\DeclareAcronym{NDSC}{short=NDSC, long=Non-Diagonal Supercell}
\DeclareAcronym{MLMD}{short=MLMD, long=machine-learning accelerated \emph{ab-initio} molecular dynamics}
\DeclareAcronym{PES}{short=PES, long=potential energy surface}
\DeclareAcronym{RMSE}{short=RMSE, long=root mean squared error}
\DeclareAcronym{FOM}{short=FOM, long=figure of merit}

\begin{document}

\title{Fine-tuning foundation models of materials interatomic potentials with frozen transfer learning}

\author{Mariia Radova}
\affiliation{Department of Chemistry, University of Warwick, Gibbet Hill Road, Coventry CV4 7AL, United Kingdom}
\affiliation{Department of Physics, University of Warwick, Gibbet Hill Road, Coventry CV4 7AL, United Kingdom}

\author{Wojciech G. Stark}%
\affiliation{Department of Chemistry, University of Warwick, Gibbet Hill Road, Coventry CV4 7AL, United Kingdom}

\author{Connor S. Allen}%
\affiliation{Department of Physics, University of Warwick, Gibbet Hill Road, Coventry CV4 7AL, United Kingdom}
\affiliation{Warwick Centre for Predictive Modelling, School of Engineering, University of Warwick, Gibbet Hill Road, Coventry CV4 7AL, United Kingdom}

\author{Reinhard J. Maurer}
\email{r.maurer@warwick.ac.uk}
\affiliation{Department of Chemistry, University of Warwick, Gibbet Hill Road, Coventry CV4 7AL, United Kingdom}
\affiliation{Department of Physics, University of Warwick, Gibbet Hill Road, Coventry CV4 7AL, United Kingdom}
\author{Albert P. Bart\'ok}
\email{apbartok@gmail.com}
\affiliation{Department of Physics, University of Warwick, Gibbet Hill Road, Coventry CV4 7AL, United Kingdom}
\affiliation{Warwick Centre for Predictive Modelling, School of Engineering, University of Warwick, Gibbet Hill Road, Coventry CV4 7AL, United Kingdom}

\date{\today}

\begin{abstract}
Machine-learned interatomic potentials are revolutionising atomistic materials simulations by providing accurate and scalable predictions within the scope covered by the training data. However, generation of an accurate and robust training data set remains a challenge, often requiring thousands of first-principles calculations to achieve high accuracy. Foundation models have started to emerge with the ambition to create universally applicable potentials across a wide range of materials. While foundation models can be robust and transferable, they do not yet achieve the accuracy required to predict reaction barriers, phase transitions, and material stability. This work demonstrates that foundation model potentials can reach chemical accuracy when fine-tuned using transfer learning with partially frozen weights and biases. For two challenging datasets on reactive chemistry at surfaces and stability and elastic properties of tertiary alloys, we show that frozen transfer learning with 10-20\% of the data (hundreds of datapoints) achieves similar accuracies to models trained from scratch (on thousands of datapoints). Moreover, we show that an equally accurate, but significantly more efficient surrogate model can be built using the transfer learned potential as the ground truth. In combination, we present a simulation workflow for machine learning potentials that improves data efficiency and computational efficiency. 


\end{abstract}

\maketitle


\section*{\label{sec:intro}Introduction}

Foundation models are large-scale machine learning models pre-trained on vast and diverse datasets that have revolutionised many domains in machine learning, enabling remarkable transferability and adaptability across various tasks.
They were first popularised in \ac{NLP} in the 2010s through models such as  BERT \cite{BERT} and GPT \cite{GPT}, then became used in computer vision, for example, in Vision Transformers \cite{vision-transformers}), and image recognition applications such as ResNet \cite{ResNet} and DenseNet \cite{DenseNet}
Recently, foundation models were generated in atomistic modelling and materials science, with a particular focus on materials discovery applications.
Examples of foundation models trained on diverse databases of chemical structures are MEGNet \cite{MEGNet}, GemNet \cite{GEMNet}, CHGNet \cite{MPtrj}, MACE-MP \cite{MACE-MP}, ALIGNN \cite{ALIGNN}, GNoME \cite{GNoME}, and more recent models: GRACE \cite{GRACE}, EquiformerV2 \cite{EquiformerV2}, MatterSim {\cite{MatterSim} and Orb \cite{Orb}.
Foundation models represent a shift towards well-generalised models as opposed to problem-specific potentials, although often at the price of reduced accuracy in predictions.
Fine-tuning foundation models for a specific task presents a data-efficient compromise, as data generation can be exceedingly costly, especially in atomistic modelling, where it relies on first principles electronic structure methods.

In the domain of atomistic modelling, foundation models typically use Graph Neural Network (GNN) architectures that aim to capture atomic interactions via message passing.
Message passing allows learning atomic representation through the exchange of messages between atoms, represented by nodes in the graph. 
The MACE~\cite{MACE} architecture incorporates many-body messages and equivariant features which facilitate capturing symmetry properties of the atomic structures. 
Recently, large foundation models based on MACE were trained using the Materials Project dataset (MPtrj)~\cite{MPtrj} that have shown impressive performance on a wide variety of benchmark systems~\cite{MACE-MP}.

However, when the dynamics of complex systems are modelled, such as gas-surface dynamics or studying phase transitions, the training database of foundation models will inevitably under-represent atomic environments relevant to achieving quantitative predictions.
In such cases, fine-tuning exploits the capability of the model to generalise, achieving excellent accuracy using a limited amount of data. 
Naive fine-tuning of MACE-MP foundation models — starting from the final checkpoint of the pre-trained model — has been shown to be data-efficient \cite{naive}.
However, this approach may lead to catastrophic forgetting and can pose risks of training instability due to continued updating of deeper network layers, which are particularly susceptible to divergence \cite{feature_distortion, catastrophic_forgetting}.
To address catastrophic forgetting, multi-head fine-tuning has recently been introduced for MACE-MP models \cite{MACE-MP}. It focuses on maintaining transferability across the systems represented in the MPtrj dataset and allows training on data obtained from multiple levels of electronic structure theory.
Another fine-tuning method has been suggested that transforms the MACE-MP descriptors into random-feature (RF) maps, focussing on data efficiency~\cite{franken}.
Frozen transfer learning has been implemented for CHGNet\cite{MPtrj}, demonstrating its robustness and accuracy, however, the aspect of data efficiency of their approach remained unexplored.
Notably, the fine-tuning of CHGNet required a substantial dataset, with more than 196,000 structures used in the fine-tuning database, demonstrating a similar performance and database requirement as those of a from-scratch CHGNet model.

In this work, we apply the transfer learning method with partially frozen weights and biases for fine-tuning \ac{MLIP} foundation models (Figure \ref{fig:fig1}a). Our aim is to use the foundation model as a stepping stone to create a tailor-made model that can describe the dynamics of a specific system as accurately as possible with as little data as possible. We discuss two challenging systems to show how transfer learning on the atomistic foundation models can be a more data-efficient way of generating highly accurate models than training them from scratch with only task-specific data: dissociative adsorption of molecular hydrogen on copper surfaces, and a ternary alloy (Figure \ref{fig:fig1}c). Furthermore, we show that the fine-tuned foundation model can be used to generate ground truth labels for a more efficient surrogate model based on the Atomic Cluster Expansion (ACE) \cite{ace}. In doing so, we benefit from the data efficiency of the fine-tuning process and can still produce a model capable of rapid inference to tackle large-scale or massively parallel simulations.

\begin{figure}
    \centering
    \includegraphics[width=1\linewidth]{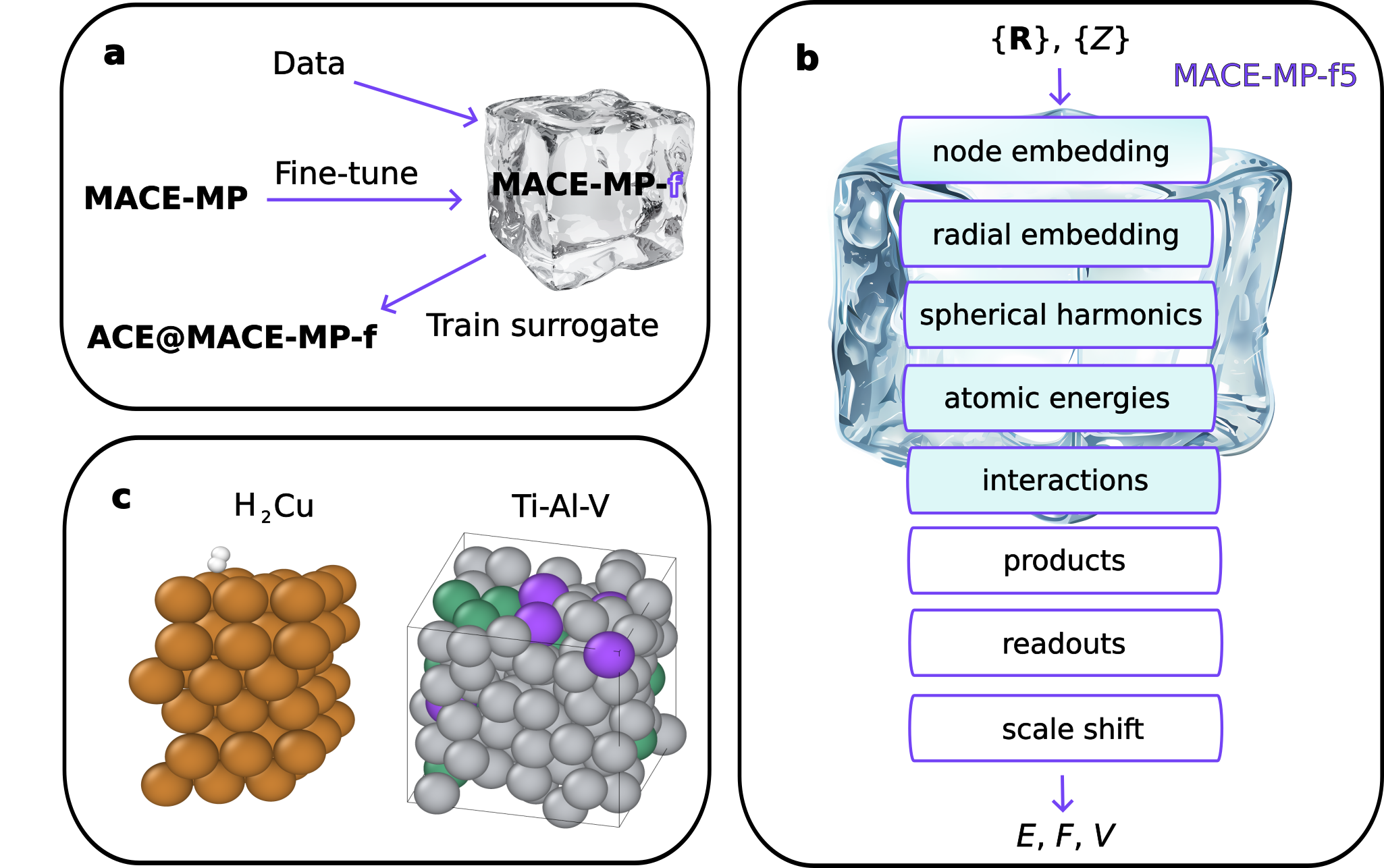}
    \caption{\textbf{A schematic representation of the key concepts of the MACE-freeze method.} (a) The workflow for training a model based on MACE-MP using transfer learning and using it to generate accurate data to fit a fast ACE surrogate model. (b) Transfer learning by freezing parameters in the model layers. (c) The atomic systems used in our benchmarks.}
    \label{fig:fig1}
\end{figure}

\section*{\label{sec:results}Results and Discussion}

\subsection*{Transfer learning}

The transfer learning technique implemented in this work for fine-tuning MACE-MP foundation models involves controlled freezing of neural network model layers.
In this process, the model parameters corresponding to a particular model layer that are kept fixed during training.
In other words, backpropagation is only carried out on the active neural network layers (Figure \ref{fig:fig1}b).
This technique has proved to be efficient in the fine-tuning of convolutional neural network models trained for image recognition~\cite{cnn_transfer3,cnn_transfer,cnn_transfer2}.
The data efficiency of transfer learned models is justified by the fact that common features or patterns deduced in the original training phase are retained.
These low-level components of the model are expected to be general, and the remaining adjustable parameters can be reliably fitted using the scarce training data provided during the transfer learning procedure.
Fixing parameters in some of the layers also reduces the training time for frozen transfer learned models compared to back-propagating the information through the whole model.
In cases when there is only little data available for training a from-scratch-trained neural network, transfer learning represents an efficient alternative, especially if generating new data is computationally expensive, for example in atomistic simulations.

We created the \verb+mace-freeze+ patch\cite{mace-freeze} to the MACE software suite that allows to freeze layers or parameter tensors in any MACE model 
to fine-tune them using a particular dataset of interest.
This approach allows for retaining the features learned from the MPtrj dataset of the MACE-MP foundational model and adapting the later layers to the new task.
The models, which we refer to as MACE-freeze, trained using \verb+mace-freeze+ patch retain the same architecture as the original MACE model used for fine-tuning and only differ in which layers or parameter groups are frozen.

\subsection*{Transfer learned model data efficiency and performance: H\textsubscript{2}/Cu}

To show the efficiency of the MACE-freeze models, we compare their accuracy to from-scratch-trained MACE models trained on the same dataset. The hyperparameters of the from-scratch-trained MACE models are fully optimised on the dataset.
We aim to demonstrate that a ``universal'' model such as MACE-MP can be fine-tuned using transfer learning to perform a specific task at least as well as a from-scratch MACE model that was trained specifically for this task only.
Moreover, we show that the transfer learned model can outperform, at least in the low-data regime, the from-scratch model in the high-data regime.
As a consequence, a much smaller number of training points are required to achieve similar accuracy of predictions with the MACE-freeze model, than the bespoke MACE model.

The ``small'', ``medium'' and ``large'' MACE-MP foundation models were used as the basis of our transfer learned models using a dataset on reactive hydrogen chemistry on various facets of copper surfaces \cite{stark_machine_2023}.
The database contains 4230 structures and was obtained using reactive gas-surface scattering molecular dynamics simulations and a committee uncertainty-driven active learning algorithm.
A from-scratch MACE model was trained using this dataset, with hyperparameters optimised previously, which were validated with k-point cross-validation and \ac{MD}.\cite{stark_benchmarking_2024}

The MACE-freeze models retain the architecture of the foundation models and use varying numbers of frozen layers of the original MACE-MP ``small'', ``medium'' and ``large'' foundation models. We show the learning curves of different ``small'' freeze models, as well as the from-scratch-trained MACE model in Fig.~\ref{fig:lc_layers}a-c.
We trained models where parameters are fixed in all layers except the readouts (MACE-MP-f6) and then incrementally we allow parameters to vary in the product layer (MACE-MP-f5), and the interaction parameters (MACE-MP-f4), while in MACE-MP-f0 all layers are active.

\begin{figure}
    \centering
    \includegraphics[width=1\linewidth]{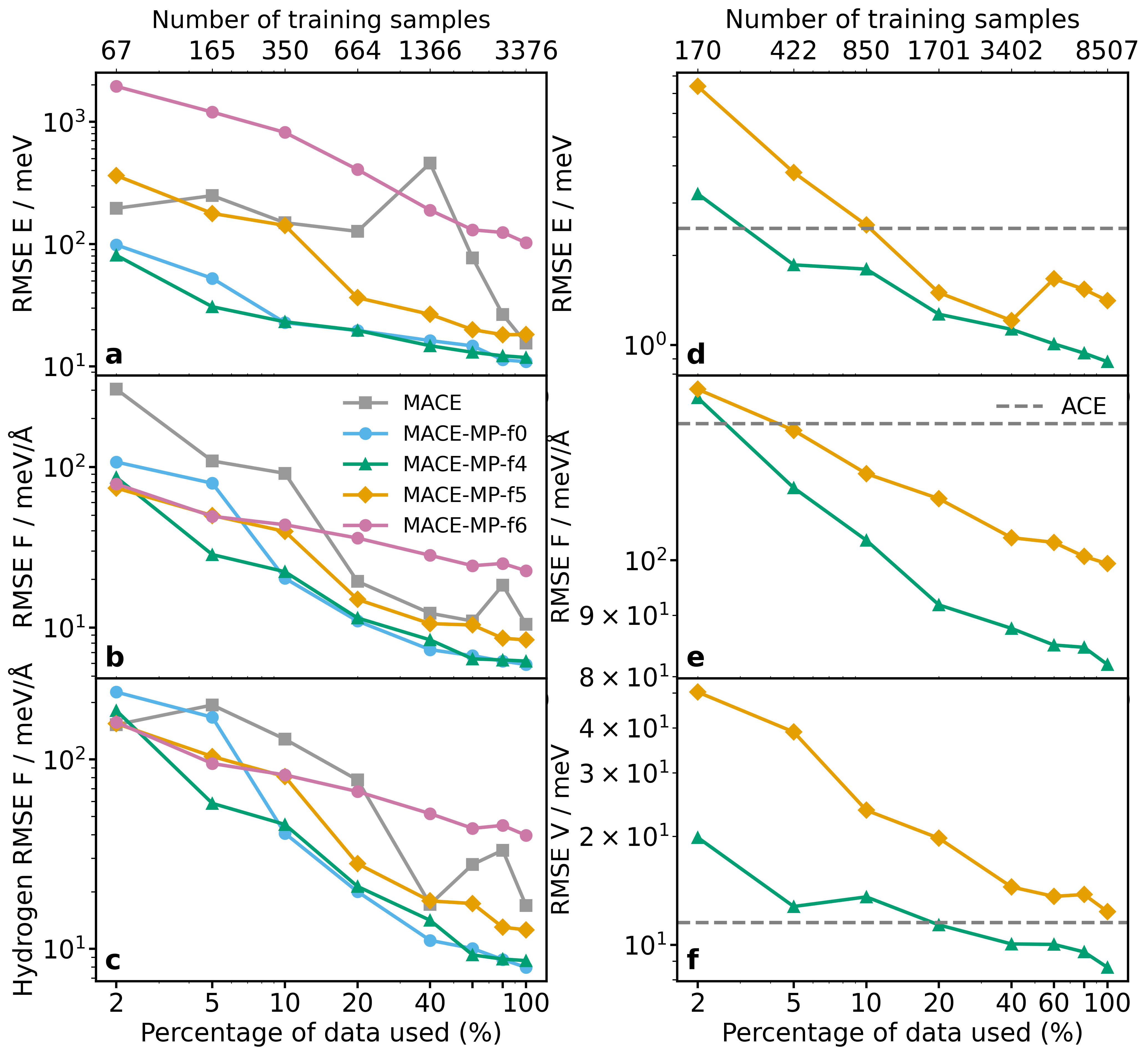}
    \caption{\textbf{Transfer learning curves for Hydrogen on Copper surface and for the Ti-Al-V alloy systems, trained based on the ``small'' foundation model.}
    For the Cu-H$_2$ system, \acp{RMSE} of energies, force components and force components of the H atoms only are shown in panels (a), (b) and (c), respectively.
    For the Ti-Al-V alloy system, \acp{RMSE} of energies, force components and virial stress components are shown in panels (d), (e) and (f), respectively.
     The points correspond to the percentages of the respective datasets, namely 2, 5, 10, 20, 40, 60, 80 and 100\% for both systems.
    The layers were frozen correspondingly to f6 (pink circles), f5 (yellow diamonds), f4 (green triangles), and f0 (blue circles).
    Grey squares mark the learning curve of the from-scratch-trained MACE model in case of the Cu-H$_2$ system, and
    the grey dashed line marks the errors of the custom ACE model in case of the Ti-Al-V alloy system.}
    \label{fig:lc_layers}
\end{figure}

In the low-data regime, where the models are trained on a small percentage of the original training datapoints, all the represented freeze models with the exception of the least flexible model, f6, perform better than the from-scratch-trained MACE model.
As the number of frozen layers decreases, the predictive performance on energies and forces improves, peaking at f4.
Allowing more flexibility by further reducing the number of frozen layers does not improve the predictive performance: the MACE-MP-f0 model, in which all parameters were allowed to update, has similar validation errors to that of MACE-MP-f4.
The reduced number of trainable parameters in MACE-MP-f4, however, have the minor added benefit of reducing the computational cost of training (Supplementary Figures S1 and S2).
Fig.~\ref{fig:lc_layers} also demonstrates that the lower layers of the network, as fitted in the original MACE-MP model do not benefit from further fine-tuning and may be reused.
The benefit of the reduced cost of training can be of particular importance for users with limited computational resources.
The superior performance of the transfer-learned models suggests that the models benefit from the pre-training due to other structures present in the MACE-MP training set, resulting in transferable and robust descriptor embeddings. 

Having found the optimum number of frozen layers to be four in our applications, we used this setting for further benchmarks.
Supplementary Figure S3 shows that the size of the foundation model does not significantly contribute to the accuracy of the transfer-learned models, but ``medium'' and ``large'' models are computationally more demanding.
For this reason, in the subsequent models, we use the ``small'' MACE-MP foundation model for our further investigations.

At 20\% of the training set (664 configurations), the MACE-MP-f4 model shows a similar level of accuracy (measured by root mean squared error, RMSE) on energies and total forces as the from-scratch-trained MACE model trained using all configurations in the training set (3376 points). Notably, MACE-MP-f4 predicts forces on hydrogen atoms that are significantly more accurate than the from-scratch-trained model hydrogen-only forces (Figure \ref{fig:lc_layers}c). Stark \textit{et al.} previously found that the force errors on hydrogen atoms predicted by a from-scratch-trained MACE model are considerably higher, which is the key limiting factor in determining the accuracy of dynamic reaction probabilities such as sticking probabilities.~\cite{stark_benchmarking_2024} 
In contrast, all the transfer-learned models have resulted in better force error measures and more balanced force RMSEs across copper and hydrogen atoms. 

\subsection*{Model validation and ACE-surrogate for H\textsubscript{2}/Cu dynamics}

To independently validate the transfer-learned MACE-MP models, we assess the prediction accuracy of the structure, reaction barriers and dynamic scattering probabilities. Supplementary Figure S4 reports the energy-volume curve of bulk Cu as predicted by \ac{DFT}, MACE-MP, MACE-MP-f4 10\%, MACE-MP-f4 20\%, and the previously published MACE model~\cite{stark_benchmarking_2024}. While the results obtained with the MACE-MP models deviate from \ac{DFT} methods, we may attribute the differences to the underlying methods and employed electronic structure packages: the PBE functional \cite{PBE} and VASP~\cite{kresse_ab_1993, kresse_efficient_1996,kresse_efficiency_1996} were used to evaluate the MP database while the SRP functional \cite{SRP} and FHI-aims~\cite{blum_ab_2009} were used to generate the Cu-H database. 
Using only 10\% of data from our previous database for transfer learning, the agreement of our MACE-MP-f4 models is in excellent agreement with \ac{DFT}, outperforming the from-scratch-trained MACE model which was fitted using the entire database.


An accurate description of lattice expansion can play a significant role in predicting reaction probabilities at metal surfaces.~\cite{mondal_thermal_2013} Thus, we explore the ability of the various approaches to capture the surface temperature-mediated lattice expansion by running NPT simulations at 9 different temperatures, between 200 and 1000~K. 
We observed that the foundation MACE-MP models compared to our other models underestimate the lattice constants by 0.035~$\textrm{\AA}$ across all the considered temperatures (Supplementary Figure S5), which can be attributed to differences in \ac{DFT} functionals and codes used in generating the respective training databases.
Given that the MACE-MP foundation models were originally trained on \ac{DFT} data obtained using the PBE functional, this behaviour is expected as with PBE it was reportedly not possible to reproduce adsorption-related experimental observables for H\textsubscript{2} at Cu surfaces.
To remedy this failure, the SRP48 functional was constructed as a combination of PBE and RPBE functionals (52\% and 48\%, respectively), to fit the experimental data, giving the most reliable prediction capabilities for this system \cite{diaz_chemically_2009, nattino_effect_2012}.
Notably, already after including 10\% of the structures from our Cu-H database the results obtained with a from-scratch-trained MACE model, trained on 100\% of the data, and MACE-MP-f4 transfer models are in very close agreement.


To assess the ability of the models to predict the reaction barriers of H\textsubscript{2} dissociation at different Cu surfaces we evaluated the \acp{MEP} using \ac{CINEB} method (Fig.~\ref{fig:meps}).
MACE-MP models were not able to predict the \acp{MEP} well for any of the surfaces we considered.
They also predicted spurious local minima that were not found with \ac{DFT} or MACE models, e.g. at Cu(111) surface around 3~$\textrm{\AA}$ or at Cu(211) at 3.5~$\textrm{\AA}$.
The spurious minima disappear in \acp{MEP} generated with both the 10\%, and 20\% transfer models, and the predictions of barriers match the reference results well.


\begin{figure}
    \centering
    \includegraphics[width=1.0\linewidth]{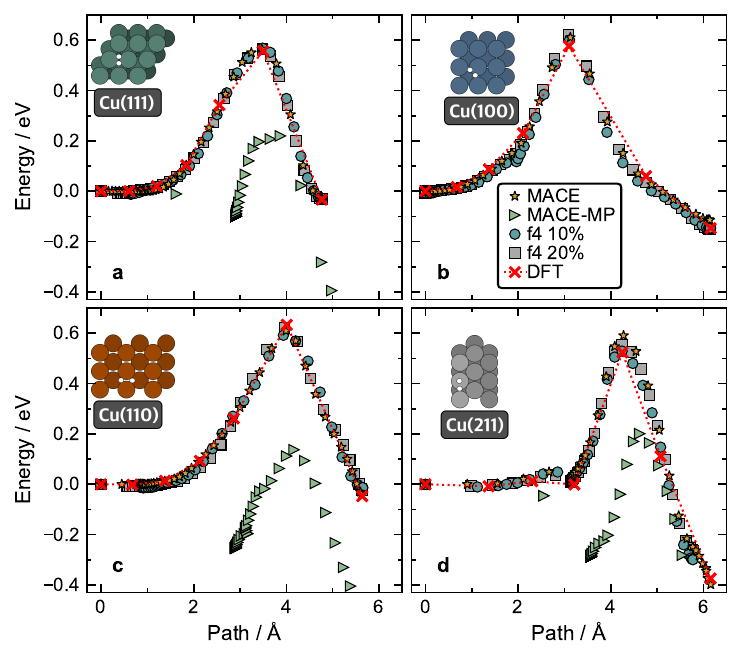}
    \caption{\textbf{Minimum energy paths obtained using \ac{CINEB} method for H\textsubscript{2} dissociative adsorption on different copper surfaces.} Potential energy values are shown along the reaction path ($\textrm{\AA}$), evaluated using \ac{DFT} (red \textbf{$\pmb{\times}$}) and the \acp{MLIP} included in our study: MACE (yellow stars), MACE-MP (green triangles), MACE-MP-f4 10\% (blue circles), and MACE-MP-f4 20\% (grey squares) for H\textsubscript{2} dissociation at Cu(111) (a), Cu(100) (b), Cu(110) (c), and Cu(211) (d) surfaces.
    The top-down views of the configurations at the transition states are included on the respective plots. All DFT data points included in the plot were taken from Ref.~\cite{stark_machine_2023}. For Cu(100), MACE-MP predictions are beyond the shown energy scale in panel b.}
    \label{fig:meps}
\end{figure}

One of the most important dynamical observables for investigating adsorption processes at metal surfaces is the sticking probability ($P_{\mathrm{stick}}$), which is the probability of an $H_2$ molecule with a given vibrational ($\mathrm{\nu}$) and rotational ($J$) initial state to dissociatively adsorb at the surface, as opposed to scattering from it. The sticking probability is calculated as the ratio between reactive events and the total number of simulated scattering events ($P_{\mathrm{stick}} = n^{\mathrm{dissoc}}_{\mathrm{traj}}/n^{\mathrm{all}}_{\mathrm{traj}}$).
Here, we evaluate the sticking probability for the adsorption of H\textsubscript{2} in the ground ($\mathrm{\nu}$=0) and first excited ($\mathrm{\nu}$=1) vibrational states and the first excited rotational state ($J=1$) at Cu(111) (925~K) and H\textsubscript{2}($\mathrm{\nu}$=1, $J=1$) at Cu(211) (925~K) with MACE-MP-f4 10\%, and compare the results with the models trained on the entire database (Fig.~\ref{fig:sticking_exp_mace}). As previously discussed, reactive scattering at 925~K surface temperature is a challenging validation of the models as the training dataset was generated from data samples drawn from low-temperature scattering, which does not guarantee that the trained models generalise to high-temperature scattering.~\cite{stark_benchmarking_2024}
The dynamics simulated with MACE-MP in all cases led to an overestimation of sticking probabilities. This overestimation is unsurprisingly largest at lower collision energies, where the prediction of the reaction barrier is the greatest determining factor in obtaining accurate sticking probabilities. As shown before in Fig.~\ref{fig:meps}, MACE-MP models severely underestimate the \acp{MEP}, leading to increased sticking probabilities.
Transfer learning on just 10\% of our database leads to a significant improvement, resulting in sticking probabilities very close to the experimental results depicted by the red line.
The agreement is comparatively good with our previous from-scratch-trained MACE model trained exclusively on our \ac{DFT}-based database.
Here, the only visible discrepancy in the predictions of the sticking probability is at higher collision energies (above 0.5~eV for H\textsubscript{2}($\mathrm{\nu}$=0), and 0.3~eV for H\textsubscript{2}($\mathrm{\nu}$=1)).
The deviation is less than 10\%, which is at a similar level to the difference between the experiment and the MACE predictions. Notably, as the experimental reference is based on permeation experiments~\cite{kaufmann_associative_2018}, the results have to be scaled to be comparable to theoretical predictions.~\cite{stark_machine_2023} As described in detail in previous work, we have scaled the experiment to match the previous theoretical predictions of the from-scratch-trained MACE model.~\cite{stark_machine_2023}
The simulation results suggest that we have reached the level of accuracy of the full-\ac{DFT} MACE model with only 10\% of structures (\ac{DFT} evaluation), thereby reducing the cost of model training by approximately a factor of ten.
The generation of training data for the Ti-Al-V system required approximately 1 million CPU hours, while the Cu-H dataset required approximately 500,000 CPU hours.

\begin{figure*}
    \centering
    \includegraphics[width=0.8\linewidth]{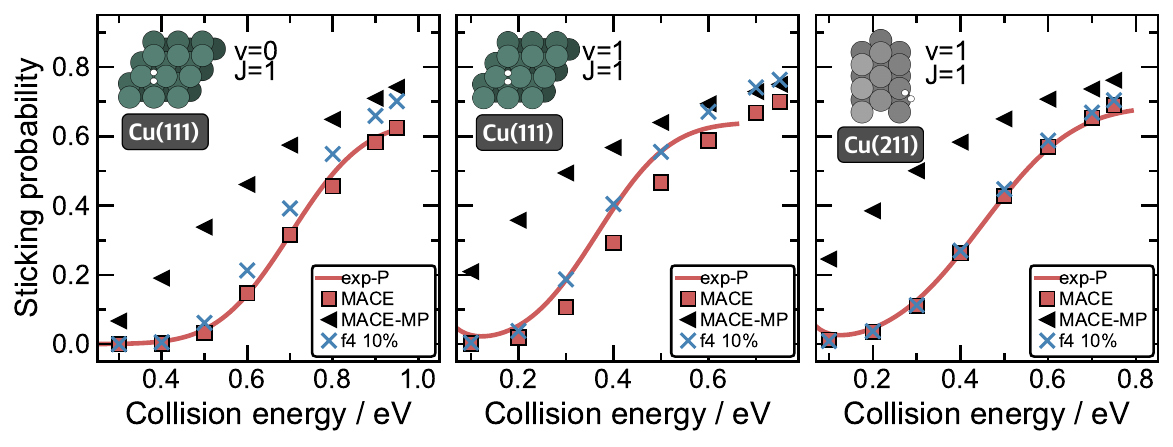}
    \caption{\textbf{Sticking probabilities for H\textsubscript{2} scattering on Cu(111) and Cu(211) at 925~K.}
    Probabilities were calculated at different collision energies using MACE (red squares), MACE-MP (black triangles), MACE-MP-f4 10\% (f4 10\%) (blue $\times$) models for the ground ($\mathrm{\nu}$=0) (left) and excited ($\mathrm{\nu}$=1) (middle) H\textsubscript{2} vibrational states at Cu(111) surface and excited ($\mathrm{\nu}$=1) state at Cu(211) ($J=1$ in every case).
    MACE refers to the model based on the \ac{DFT}-based database from Ref.~\cite{stark_machine_2023}.
    The red line represents a sticking probability obtained from the experimental results of  Kaufmann~\textit{et~al.}~\cite{kaufmann_associative_2018} (exp-P) at 923$\pm$3~K, scaled to match theoretical probabilities from Ref.~\cite{stark_machine_2023} at the highest incidence energy (saturation parameter A=0.64 for both Cu(111) sticking functions, and A=0.66 for Cu(211)).}
    \label{fig:sticking_exp_mace}
\end{figure*}

While the transfer-learned MACE-MP-f4 model provides highly accurate predictions, the time-to-solution for energy and force predictions is significantly slower than for the from-scratch-trained MACE model. The latter is a highly optimised and small model tailored for the given dataset (5~\AA{} cutoff, correlation order 2, 16x0e model size), whereas MACE-MP ``small'' has larger cutoffs and more parameters (6~\AA{} cutoff, correlation order 3, 128x0e model size.
The hyperparameters of the reference from-scratch model were justified by Stark et al. (Ref.~\cite{stark_benchmarking_2024}), who reported that increasing the model size and complexity yielded minimal improvement in the predictive performance of the from-scratch model.
As the dynamics simulations require tens of thousands of trajectories at various incidence energies to provide converged sticking probabilities, the transfer-learned MACE-MP model, despite its accuracy, comes at an almost prohibitive computational cost (Table \ref{tab:h2cu_mod_perf}). Whereas the from-scratch-trained model evaluates energies and forces per geometry in 60~ms, MACE-MP-f requires 390.5~ms on an AMD EPYC 7742 (Rome) CPU processor core. 

To address this, we take MACE-MP-f4 predictions as ground truth labels and construct a new model using the ACEpotentials.jl~\cite{witt_acepotentialsjl_2023} package, which fits linear \ac{ACE} potentials~\cite{kovacs_linear_2021}, enabling significantly faster force evaluations~\cite{stark_benchmarking_2024} than the more complex MACE-MP models.
We will refer to this model as ACE-f 10\%. Previously, optimised \ac{ACE} models trained on the full database were not able to provide sufficiently low force errors and accurate sticking probability predictions.~\cite{stark_benchmarking_2024}
We utilise data generated with MACE-freeze 10\% potential exclusively to construct a new training set for ACE-f 10\%.
This was carried out by generating a total of 600 trajectories, comprised of 5 trajectories each for H\textsubscript{2} scattering at 4 different Cu surfaces, namely, (111), (110), (100), and (211), at 3 surface temperatures (300, 600, and 900~K), at two rovibrational states (H\textsubscript{2}($\mathrm{\nu}$=0, $J=0$) and H\textsubscript{2}($\mathrm{\nu}$=1, $J=1$)), at 5 different collision energies. Such comprehensive sampling of training data would not be possible with on-the-fly ab-initio dynamics while active learning requires many more simulations and training/learning loops.~\cite{stark_machine_2023}
For each trajectory, structures were saved at the interval of 1~fs.
Then the k-means clustering method was used to choose 2000 of the most diverse configurations.
Furthermore, we included the \acp{MEP} obtained from \ac{NEB} calculations, 50 structures along reaction paths at each studied surface.
The final database contained 2200 structures.
\ac{ACE} (ACE-f 10\%) models were trained based on this database employing the hyperparameter settings used for the same system in our previous study~\cite{stark_benchmarking_2024}.
The final ACE-f 10\% model achieves excellent accuracy, which even combined with the MACE-MP-f4 10\% evaluation errors reaches chemical accuracy ( Tab.~\ref{tab:h2cu_mod_perf}).
This is a significant improvement compared to the previous  \ac{ACE} model trained directly on the full \ac{DFT}  database. The previous \ac{ACE} model was unable to reach the accuracy of the neural network-based methods~\cite{stark_benchmarking_2024}, especially for forces on hydrogen atoms, where the best performance was limited to an MAE (mean absolute error) 28.5~meV/$\textrm{\AA}$, and RMSE of 47.6~meV/$\textrm{\AA}$).
These errors are approximately 3 times larger than for ACE-f 10\% (Tab.~\ref{tab:h2cu_mod_perf}).
We attribute this discrepancy to the presence of outliers in our previous database, which was adaptively sampled using less accurate and unstable SchNet models~\cite{stark_machine_2023}.
Thus, as previously found by others \cite{Morrow2022}, constructing a large synthetic database directly using a model of high accuracy and smoothness is an efficient approach to training accurate linear models.


\begin{table}[]
    \centering
    \caption{Performance of MACE-MP-f4 10\%, ACE-f 10\%, and from-scratch (f-s) ACE and MACE models. Energy (E) and force (F) errors are included in meV and meV/$\textrm{\AA}$, respectively.
    Evaluation times (t$_{\mathrm{eval}}$) are in ms and were calculated on a single AMD EPYC 7742 (Rome) 2.25 GHz CPU processor core.} 
    \begin{tabular}{l|c|c|c|c}
    \textbf{Property} & \textbf{MACE-MP-f} & \textbf{ACE-f} & \textbf{ACE}  & \textbf{MACE}\\
     & \textbf{10\%} & \textbf{10\%} &  \textbf{(f-s)} &  \textbf{(f-s)} \\
    \hline
    \textbf{MAE (E)} & 6.7 & 4.8 & 8.6 & 11.3  \\
    \textbf{RMSE (E)} & 13.5 & 6.4 & 13.7 & 15.8 \\
    \textbf{MAE (F)} & 3.1 & 7.2 & 10.4 & 8.1 \\
    \textbf{RMSE (F)} & 10.1 & 10.9 & 18.2 & 12.9 \\
    \textbf{MAE (H-atom F)} & 4.3 & 9.0 & 28.5 & 17.7 \\
    \textbf{RMSE (H-atom F)} & 17.1 & 17.7 & 47.6 & 32.4 \\
    \textbf{t$_{\mathrm{eval}}$ (E)} & 372.6 & 9.6 & 9.1 & 60.0 \\
    \textbf{t$_{\mathrm{eval}}$ (F)} & 390.5 & 22.6 & 20.7 & 60.0 \\
    \end{tabular}
	\label{tab:h2cu_mod_perf}
\end{table}

To examine the ability of the ACE-f 10\% model to predict the actual dynamical observables, we evaluated sticking probabilities for the same systems and settings as for MACE-freeze 10\% model (Fig.~\ref{fig:sticking_exp_ace}).
In all cases, the agreement between ACE-f 10\% and MACE-freeze model is good.
Additionally, we evaluated sticking probabilities with the ACE model trained, using the same settings, on our previous database (ACE-S).
The probabilities obtained with ACE-S model match the probabilities obtained with the ACE-f 10\% model well for both rovibrational states of Cu(111), however, this is not the case for the dynamics at Cu(211), where the ACE-S model significantly underestimates, by more than 10\%, the sticking probabilities for collision energies above 0.3~eV.

\begin{figure*}
    \centering
    \includegraphics[width=0.8\linewidth]{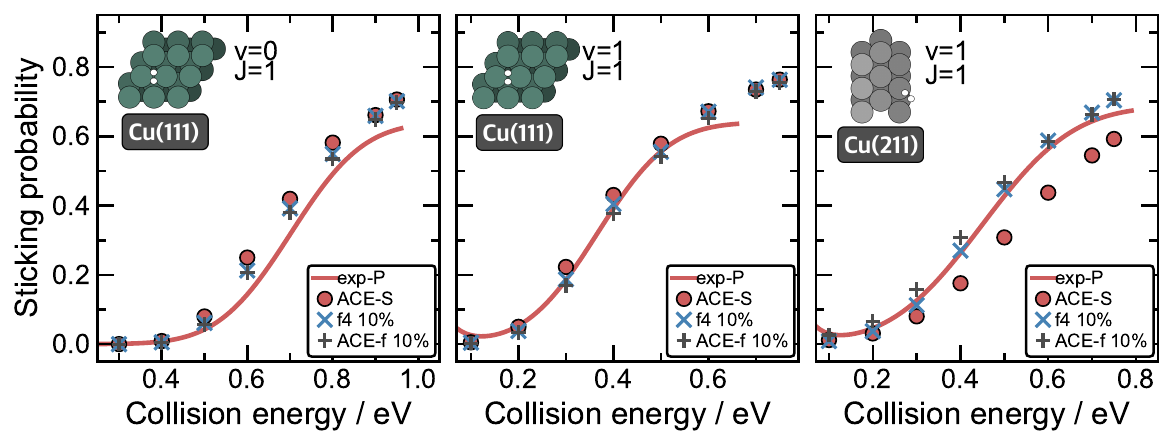}
    \caption{\textbf{Sticking probabilities for H\textsubscript{2} scattering on Cu(111) and Cu(211) at 925~K.} Probabilities were calculated at different collision energies using ACE-S (red circles), MACE-MP-f4 10\% (f4 10\%) (blue $\times$), and ACE-f 10\% (grey $+$) models for the ground ($\mathrm{\nu}$=0) (left) and excited ($\mathrm{\nu}$=1) (middle) H\textsubscript{2} vibrational states at Cu(111) surface and excited ($\mathrm{\nu}$=1) state at Cu(211) ($J=1$ in every case). ACE-S refers to the model based on the \ac{DFT}-based database from Ref.~\cite{stark_machine_2023}.
    The red line represents a sticking function obtained from the experimental results of  Kaufmann~\textit{et~al.}~\cite{kaufmann_associative_2018} (exp-P) at 923$\pm$3~K, scaled to match theoretical probabilities from Ref.~\cite{stark_machine_2023} at the highest incidence energy (saturation parameter A=0.64 for both Cu(111) sticking functions, and A=0.66 for Cu(211)).}
    \label{fig:sticking_exp_ace}
\end{figure*}

Evaluating probabilistically determined dynamical observables, such as sticking probability requires simulating hundreds of thousands of \ac{MD} trajectories, which leads to high computational expenses.
Employing \acp{MLIP} instead of traditional methods, such as \ac{DFT}, reduces that time significantly.
Due to the high generalizability of MACE-MP models, their complexity needs to be higher than \acp{MLIP} generated for a specific system, meaning the evaluation times are significantly higher.
Training fast, small models based on specialized MACE-MP models, such as MACE-MP-f models, can help mitigate this issue.
Here, a single force evaluation on an AMD EPYC7742 2.25 GHz CPU processor with our MACE-MP-f4 10\% model takes approximately 390~ms, however, by training an ACE-f 10\% model we reduce this time by more than 17 times (22.6~ms, Tab.~\ref{tab:h2cu_mod_perf}), while preserving the prediction quality, as shown in Fig.~\ref{fig:sticking_exp_ace}. 

\subsection*{Ti-Al-V data and benchmarks}

A dataset was constructed for the purpose of building \acp{MLIP} to accurately model the crystalline and liquid phases of Ti-6Al-4V (Ti 90 wt\%, Al 6 wt\%, V 4 wt\%) alloy up to 30 GPa. To model crystalline Ti-6Al-4V, the three physically observable phases below 30 GPa were considered: $\alpha$ (hcp, P$6_3$/mmc), $\beta$ (bcc, Im-3m) and $\omega$-Ti (hexagonal, P6/mmm).

To benchmark this dataset, we considered a validation set, elastic properties and vibrational properties.
The validation set consisted of a series of large simulation cells that resemble the Ti-6Al-4V stoichiometry, where developed models can be compared against the configurational energy, forces and virial stresses. 
We also calculated the elastic constants for simulation cells representative of Ti-6Al-4V in each crystalline phase using \verb|matscipy|\cite{Grigorev2024}.
Due to the dilute amount of Al and V in Ti-6Al-4V, it is not tractable to compute phonon properties in a cell representative of this stoichiometry.
Therefore, to characterise vibrational properties we consider phonon dispersion, density of states and quasi-harmonic free energy calculations of simulation cells that represent minor alloying component nearest neighbour interactions, even though the compositions correspond to significantly higher Al and V concentrations than that of Ti-6Al-4V.
Calculations were performed in 8 ($\alpha$ and $\beta$) and 12 ($\omega$) atom simulation cells, with a vibrational Brillouin Zone grid sampling of $2\times2\times2$. 

The reader is referred to Ref.~\cite{allen2025multiphasedatasettiti6al4v} where the full details of the database construction, benchmarks, and alternative \ac{MLIP} developments are presented.
Also discussed in Ref.~\cite{allen2025multiphasedatasettiti6al4v} is the development of the \ac{ACE} model for Ti-6Al-4V which is presented here as a baseline model alongside the MACE models developed in this work.
In this work, we show that frozen transfer learning, using as little as 10\% of the database,  is able to achieve superior predictive performance on our benchmarks compared to \ac{ACE} potentials fitted using the entire database. 

The ``small'' foundation model was fine-tuned using the \verb+mace-freeze+ patch on different percentages of the training database, amounting to 8507 structures in total.
We show in Fig.~\ref{fig:lc_layers}, that the validation error metrics of the MACE-MP-f4 model start to surpass the ACE baseline figures at less than 5\% of training data when considering energy and force values, and at 20\% of training data for virial stress components.
Compared to ACE, MACE-MP-f5 becomes more accurate at around 10\% data for energies, and 5\% for forces.
We note that the validation accuracy of MACE-MP-f5 on the virial stress components approaches that of the ACE model at 40\% training data, but never surpasses it.

We have applied transfer learning to all the ``small'', ``medium'' and ``large'' tiers of the MACE-MP foundation models.
We observed that, similarly to the copper-hydrogen system, the validation accuracy figures are very similar (Supplementary Figure S3) across the differently sized models.
We note that with the validation accuracy of the ``large'' model being the worst, it may indicate a small degree of overfitting.
For computational efficiency, we have used the transfer learned model based on the ``small'' MACE-MP model for our further benchmarking experiments.

The transfer models along with the \ac{ACE} model fitted on the entire database were evaluated using a set of benchmark calculations, designed to explore the accuracy of the \ac{PES} of the Ti-Al-V system at various compositions in the relevant crystalline phases.
Using \ac{DFT} calculations as our reference, elastic and vibrational properties, as well as \ac{RMSE} values of energies, forces and virial stress components in a set of independent configurations were predicted with our models. 
To compare our models, we define a \ac{FOM} to compress the error metrics into a single scale.
For each benchmark property $p$ that was calculated using model $m$ we first calculated the absolute error $\Delta_{p,m}$ which was transformed to
\begin{equation*}
    \Tilde{\Delta}_{p,m} = \frac{\Delta_{p,m}-\Delta_{p,\textrm{worst}}}
    {\Delta_{p,\textrm{best}}-\Delta_{p,\textrm{worst}}}
\end{equation*}
and used to determine the average \ac{FOM} as
\begin{equation*}
    \textrm{\ac{FOM}} = \frac{1}{N_p}\sum_p \Tilde{\Delta}_{p,m}
\end{equation*}
We note that the \ac{FOM} value of a model may vary depending on the other models included in the comparison.

We present all the metrics considered for benchmarking the Ti-6Al-4V models in Supplementary Figures S6-S9 in the SI.
Various benchmarks grouped together into four distinct categories (``validation'', ``training'', ``elastics'' and ``vibrational'') are shown in Fig.~\ref{fig:foms4}, showing the performance of transfer learned models as a function of increasing dataset size.
We collated individual benchmarks into their respective categories to demonstrate how balanced the performance of the various models is, although note that the uniform weighting of the $\Tilde{\Delta}_{p,m}$ values is an arbitrary choice.
Within the ``training'' and ``validation"" categories, we consider the \acp{RMSE} of the predicted energy, forces and virial stress quantities, compared to \ac{DFT} reference values evaluated on crystalline and liquid configurations.
Within the ``elastics'' category, we consider the absolute difference in elastic constants between our models and \ac{DFT} across the physically relevant crystalline phases, characterising second derivatives of the \ac{PES} with respect to deformations of the lattice.
Finally, we characterise vibrational properties by calculating the \ac{RMSE} in phonon dispersions, as well as considering the absolute error differences in the quasi-harmonic free energy values evaluated at 0 and 2000~K, thus providing insight into the reproduction of the force constant matrices of our models and their finite-temperature behaviour.

\begin{figure}
    \centering
    \includegraphics[width=0.9\linewidth]{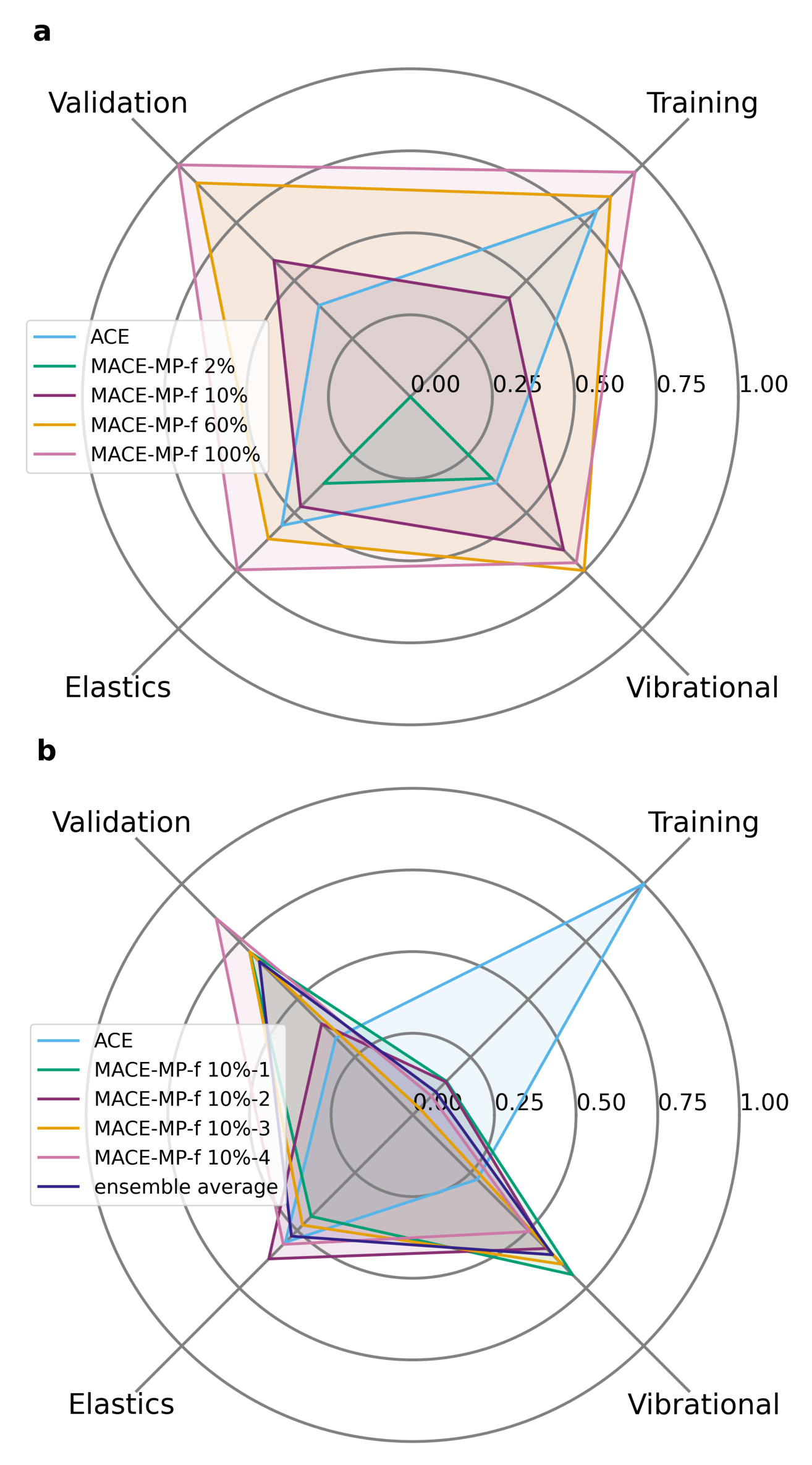}
    \caption{\textbf{Figures of merit (FOM) for f4 models relative to the custom ACE trained on the Ti-Al-V data.}. Panel (a) shows how the FOMs of the transfer models (f4) improve as the size of the database increases.
    Panel (b) displays the FOMs of an ensemble of f4 10 \% models, quantifying the uncertainty obtained via random sampling of the training set.} 
    \label{fig:foms4}
\end{figure}

We compare the performance of the transfer learned models to the baseline \ac{ACE} model trained on 100\% of the data, with the aim of achieving the same or better accuracy with as little data as possible.
We have demonstrated that transfer learned potentials based on the ``small'' MACE-MP model at the f4 (Fig.\ref{fig:foms4}) and f5 level (Supplementary Figure S10) achieved comparable performance to the \ac{ACE} model in the ``validation'', ``elastics'' and ``vibrational'' categories using only 10\% or 20\% of the database, respectively.
It is important to note that even though the Ti-6Al-4V database contains 8507 individual configurations, the diversity of the data means that some compositions in the ordered phases are only represented by a few structures. In those cases, we ensured that the structures are included in the reduced database.
To assess the uncertainty due to downsampling the original database, we trained a committee of 5 f4 models using 5 different random samples containing 10\% of the original configurations.
Using our \ac{FOM} metrics, we present the performance of the committee members and that of their ensemble average compared to the \ac{ACE} model trained on the entire database.
While there is considerable variation between the committee members, their performance on the ``validation'', ``elastic'' and ``vibrational'' benchmark groups are similar to or better than that of the \ac{ACE} model. The fact that these committee members are better than ACE indicates that transfer learning fills the gaps in the Ti-Al-V database which the database leaves out.

To demonstrate the performance of our transfer learned models, we present phonon dispersions along high symmetry lines and the phonon density of states for the relevant $\alpha$, $\beta$ and $\omega$ phases of pure titanium (Fig.~\ref{fig:ti-phonons}).
Excellent qualitative agreement with the \ac{DFT} reference can be observed at using 5\% of the database, which improves consistently as we increase the size of the database.
\begin{figure}
    \centering
    \includegraphics[width=1\linewidth]{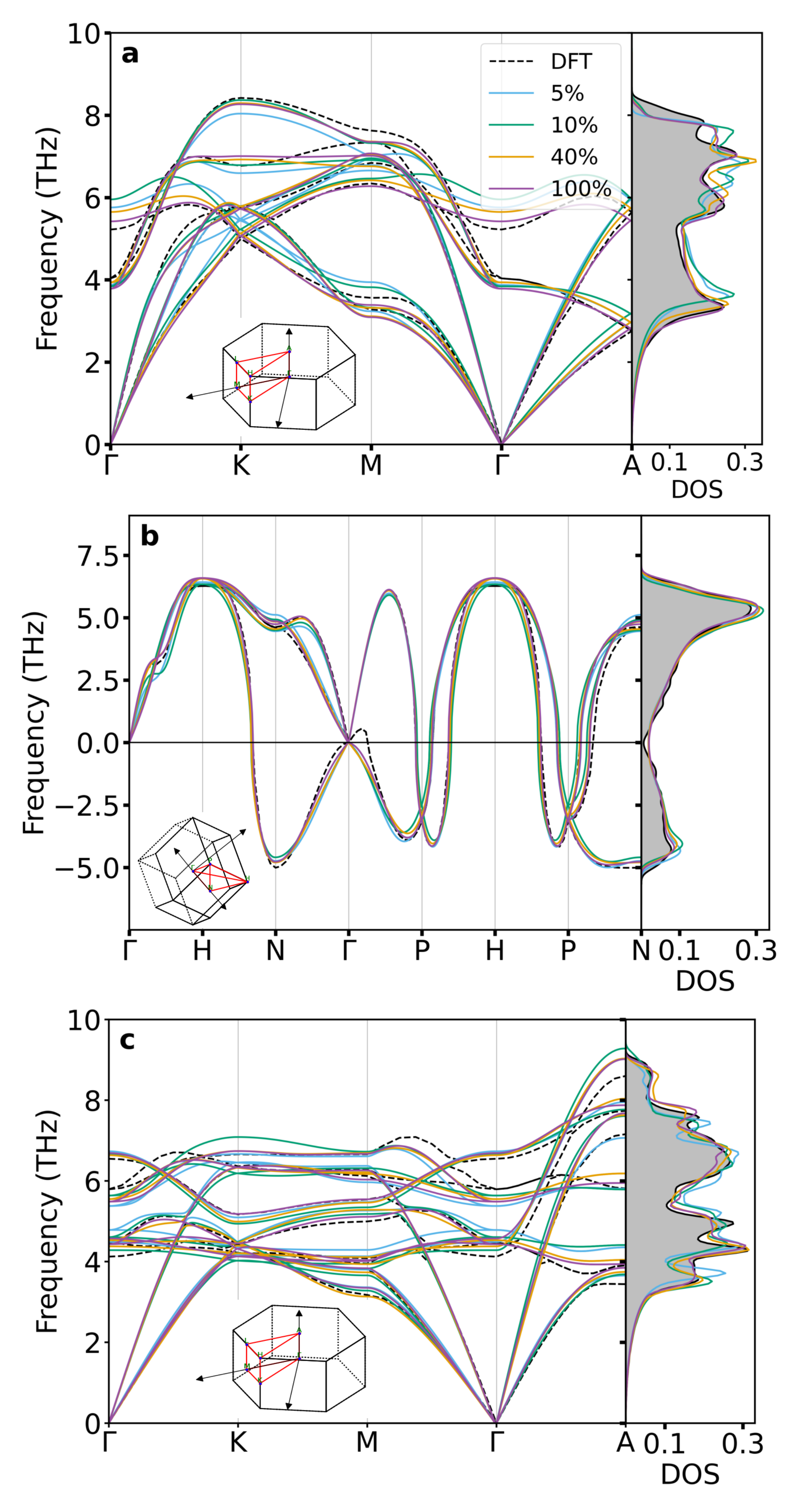}
    \caption{\textbf{Calculated phonon dispersions for (a) \textbf{$\alpha$-Ti}, (b) \textbf{$\beta$-Ti}, and (c) \textbf{$\omega$-Ti}}. Each subplot compares the predicted phonon dispersions from the f4 ``small'' models (5\% - blue, 10\% - green, 40\% - yellow, 100\% - purple) against \ac{DFT} (dashed lines and gray density of states (DOS) projection).}
    \label{fig:ti-phonons}
\end{figure}

We also compared the f4 models to from-scratch MACE models (Supplementary Figure S11), using the improvement on the overall \ac{FOM} relative to the \ac{ACE} model.
To facilitate a practical comparison, our from-scratch MACE models were trained with the same hyperparameters as the ``small'' MACE-MP foundation model.~\cite{MACE-MP}. The results show that the accuracy of the from-scratch Ti-Al-V MACE model is poorer in the low-data regime than that of the MACE-f4 Ti-Al-V model, while they perform similarly if more than 40\% of the data is used in training. 
Even in the higher training data regime the transfer models have the advantage that fitting them is computationally less expensive, as they have fewer adjustable parameters and require fewer epochs in training.
This reaffirms that the transfer models are able to provide significant benefits over from-scratch-trained models where little data is available.

\subsection*{Comparison to $\Delta$-learning}
Finally, $\Delta$ machine learning represents a commonly employed alternative to transfer learning to exploit correlations between distinct sources of data and to utilise multi-fidelity datasets simultaneously.
We compare $\Delta$ learning to frozen transfer learning by training MACE models to fit the difference between the MACE-MP predictions, regarded as baseline, and the \emph{ab initio} target using the Cu-H dataset.
To evaluate the performance of the $\Delta$-learning approach, we compared learning curves of the $\Delta$-models to that of a MACE model trained from scratch (Supplementary Figure S12). We find that the from-scratch model remains superior for all but the smallest training set sizes, where the achieved prediction errors are insufficient for production simulations.
While it may be possible to improve the performance of the $\Delta$-model by using a more complex MACE architecture, it would result in a more computationally expensive potential.
Overall, transfer learning in our tests achieved a significantly superior accuracy across low and high data regimes.

\section*{\label{sec:conclusions}Discussion}

Herein, we find that transfer learning of foundation models is a viable strategy to train \acp{MLIP} with a small amount of training data, while achieving the accuracy of from-scratch-trained potentials trained with a significantly larger number of atomic configurations.
By fixing, or freezing, parameter groups of existing foundation models the complexity of the fitting procedure is reduced.
Benchmarking our MACE-freeze approach demonstrated that the transfer learned models benefit from retaining the low-level layers in the MACE-MP neural network,  which are often interpreted as descriptors \cite{Nigam2022} \cite{e3}, and generalised from the extensive MP database.
We have shown that transfer learned models can not only achieve the accuracy of the from-scratch-trained models, but can surpass them in the accuracy of the fit and, subsequently, the predictions, sometimes able to reach the \ac{DFT} noise level of \acp{RMSE}. It is also worth pointing out that the need for hyperparameter optimisation is reduced to the weights of the observables (energies, forces, stresses) only. Other hyperparameters are inherent to the foundation model, and do not need optimising.
Transfer learned models based on MACE foundation models can be used in place of the first-principle methods such as \ac{DFT} for data generation in active learning algorithms.
The challenge of the accurate, but large models leading to slow, inefficient \ac{MD} simulations can be solved by training an efficient surrogate model, based on the potentials of the large model, as demonstrated in this work.

Our work provides a workflow that allows rapid development of fast and highly accurate tailormade \acp{MLIP} using a minimal amount of \emph{ab initio} reference data, by leveraging the information in foundation models and is, in principle, independent of the specific architecture of the foundation model. Having said that, further data efficiency and accuracy improvements may be possible if only specific types of parameter groups are frozen rather than whole layers of parameters. This will be a topic of future investigations. 

We note that there are other viable transfer learning approaches to exploit scarce amounts of \emph{ab initio} data to train accurate \acp{MLIP}.
For example, Gardner et al. suggested training, from scratch, NequIP models based on atomic configurations obtained from simulations using \acp{MLIP} and using transfer learning to refine the resulting models with a small amount of \ac{DFT} labels.\cite{Gardner}
Such a targeted approach could be very \emph{ab initio} data-efficient, as the original model is specifically adapted to the configurational space of interest.
However, as discussed by Gardner et al., ``the synthetic source matters'', which in most cases means a good quality \ac{MLIP} is a prerequisite to generating relevant atomic configurations.
Our frozen learning relies on a much more general foundational model and achieves specificity by fixing parameter groups in the refining step, thereby avoiding the need to generate synthetic data and train a from-scratch model to be transfer-learned.

Frozen transfer learning inevitably reduces the transferability that foundation models provide in favour of creating a model suitable for a narrow application domain. The extent of this scope-narrowing requires further investigations. Other techniques, such as multi-head learning, have been proposed to retain transferability by replaying the original data on which the foundation model has been trained. Such approaches could be combined with frozen transfer learning in the future to retain a higher degree of transferability.

Negative transfer, where knowledge from a source domain impedes rather than enhances performance in the target domain, can pose a challenge in transfer learning \cite{negative_transfer}.
This phenomenon often arises when the source and target distributions exhibit large discrepancies, causing the transferred features to introduce biases that hinder model convergence.
However, when the pre-trained model has been trained on a sufficiently diverse and representative dataset that captures the chemical and structural diversity relevant to the target task, the learned embeddings and interaction patterns are more likely to generalise effectively \cite{transfer_strategies}.
Therefore, selecting a relevant foundation model for the target task is important to minimise the risk of negative transfer.
For instance, if one is interested in inorganic materials domain, using a foundation model trained on organic molecules such as MACE-OFF may cause negative transfer effects.
Supplementary Figure S13 illustrates the spatial relationships between two specific training datasets and a sampled subset of MPtrj data, as interpreted through the learned features.
It is apparent that both datasets exhibit overlap with the MPtrj dataset; however, the Ti-Al-V dataset demonstrates a notably smaller degree of overlap with the foundation model data compared to the H$_2$/Cu dataset.
Furthermore, the Ti-Al-V dataset occupies a substantially larger feature space than the H$_2$/Cu dataset.
We observed that the Ti-Al-V MACE-MP-f4 model converged more rapidly than the MACE-MP-f5 model.
This may indicate that as the representation of a given system diverges further from that captured by the foundation model, more flexibility is required to achieve an accurate fit.
In our context, a greater number of parameters may need to remain unfrozen. We have shown that negative transfer did not occur even in this case.
However, fully leveraging the strengths of the frozen transfer learning method relies on a key assumption: that the data of interest is well-represented at the lower-level features learned by the foundation model.
In other words, that the data of interest is not too far from the foundation model training set.

\section*{\label{sec:methods}Methods}
\subsection*{Model hyperparameters}

 \textbf{Hydrogen on Copper surface transfer models}. 
 The following hyperparameters were used for training on this dataset for all frozen transfer learning models based on the ``small'', ``medium'', and ``large'' MACE-MP foundation models: a learning rate of 0.01, a cutoff distance of 6~Å, and a batch size of 16. The force and energy weights were set to 100 and 1, respectively, for the first 1200 epochs, and 1000 and 100, respectively, for the last 300 epochs. The models trained for a total of 1500 epochs. The freeze parameter was set to 0, 3, 4, 5, or 6. We note that training for such a large number of epochs was not necessary, but here and for other models, it was done to ensure convergence and stability across the models trained on different subset sizes.
 \newline
 \newline
 \textbf{Hydrogen on Copper surface delta models}. 
The delta learning models used for refining the ``small'', ``medium'' and ``large'' MACE-MP foundation models had the following hyperparameters: a correlation order of 2, a cutoff distance of 4 Å, a model size of 16×0e, and 2 interaction layers.
Notably, increasing the cutoff distance to 6 Å has shown no improvement in the on the test set RMSE.
The initial loss function had energy and force weightings of 1 and 100, respectively, with the energy weighting increasing to 1000 after 1200 epochs. The models were trained for a total of 1500 epochs.
\newline
\newline
\textbf{Ti-Al-V transfer models}.
The ``small'' MACE-MP foundation model was fine-tuned for this task using a learning rate of 0.01, a cutoff distance of 6~Å, and a batch size of 32. The force and energy weightings were both set to 100, with stochastic weight averaging (SWA) starting at 1200 epochs, using SWA force and energy weightings of 10 and 1000, respectively. The freeze parameter was set to 3, 4, or 5. The maximum number of epochs was 2000, with SWA starting at 1500 epochs, for consistency across models trained on different subset sizes.
\newline
\newline
\textbf{ACE surrogate model}.
The models were trained with ACEpotentials (\url{https://github.com/wgst/ACEpotentials.jl})~\cite{witt_acepotentialsjl_2023}, version 0.8.2. The cutoff distance was set to 6~Å. A correlation order of 4 was used, with polynomial degrees of 18, 12, 10, and 8, respectively. The energy and force loss function weightings were set to 0.54 and 0.01, respectively.

\subsection*{Ti-Al-V dataset}

The underlying Ti-Al-V data was obtained from \ac{DFT} calculations with the plane wave package \verb|CASTEP|(v24.1)\cite{clark_first_2005}.
On-the-fly ultrasoft pseudopotentials were generated for Al, V and Ti with respective valence electronic structures: 3s$^2$3p$^1$, 3s$^2$3p$^6$3d$^2$4s$^2$, and 3s$^2$3p$^6$3d$^2$3s$^2$.
The PBE~\cite{PBE} level of theory was used to model exchange-correlation.
The parameters for our \ac{DFT} calculations are configured to ensure convergence to within sub-meV per atom, compared to an excessive basis set and k-point sampling.
This convergence criterion was met by using a plane wave energy cutoff of 800~eV and by sampling the electronic Brillouin Zone with a Monkhorst-Pack grid spacing of 0.02~\AA$^{-1}$. 

To characterise potential local ordering in each crystalline phase ($\alpha$ (hcp, P$6_3$/mmc), $\beta$ (bcc, Im-3m) and $\omega$-Ti (hexagonal, P6/mmm)), the \ac{NDSC} method \cite{lloyd-williams_lattice_2015,allen_optimal_2022} was extended as a data reduction tool to efficiently sample atomic species ordering.
In this method, a series of \acp{NDSC} are generated for pure Ti in each crystalline symmetry, from which atomic species are randomly swapped from Ti to Al and V.
The simulation cells are then volumetrically scaled and randomly deformed, with atomic positions being perturbed according to a normal distribution with a standard deviation of $0.10$~\AA.

To characterise the liquid phase of Ti-6Al-4V, we utilise the \ac{MLMD} feature in \verb|CASTEP| by Stenczel \textit{et al} \cite{stenczel_machine-learned_2023}.
In \ac{MLMD}, molecular dynamics simulations are performed using a combination of \ac{DFT} and on-the-fly generated \acp{MLIP} to propagate the dynamics, whereby a decision making algorithm is used to switch between the \ac{DFT} and \ac{MLIP} calculator, which is constantly updated with \ac{DFT} datapoints.
The ultimate result of switching between \ac{DFT} and the surrogate \ac{MLIP} is that in a given simulation one may consider a much larger number of timesteps in a set amount of computation time.
\ac{MLMD} was performed on simulation cells containing 54 and 128 atoms, where the stoichiometry resembled Ti-6Al-4V as close as possible. For full details, the reader is again referred to Ref.~\cite{allen2025multiphasedatasettiti6al4v}.

\subsection*{Simulation details}

Molecular dynamics simulations (excluding NVT and NPT simulations) for H\textsubscript{2}/Cu were run using NQCDynamics~\cite{gardner_nqcdynamicsjl_2022} package (\url{https://github.com/NQCD/NQCDynamics.jl}, version 0.14.0). NPT and NVT (Langevin MD) simulations, as well as NEBs, were evaluated using the Atomic Simulation Environment~\cite{larsen_atomic_2017} (\url{https://gitlab.com/ase/ase}, version 3.23.0).

Initial structures for simulations included structures in which the hydrogen molecule is situated 7~{\AA} above the copper surface. The initial vibrational and rotational states were established using the Einstein–Brillouin–Keller (EBK) method~\cite{larkoski_numerical_2006}, implemented within NQCDynamics. Polar and azimuthal angles were chosen randomly. Initial positions and velocities of surface atoms were established by running Langevin molecular dynamics at set temperatures using adequate lattice constants, based on NPT simulations, as detailed in Ref.~\cite{stark_benchmarking_2024}. 
Sticking probabilities were evaluated using data from 10,000 H\textsubscript{2}/Cu trajectories (at every collision energy, surface facet and rovibrational state). The maximum simulation time was set to 3~ps with a time step of 0.1~fs. However, the trajectories were stopped when the following conditions were met: the distance between adsorbed hydrogen atoms was larger than 2.25~{\AA} (dissociation), or the distance between hydrogen molecule and surface exceeded 7.1~{\AA} (scattering). Sticking probabilities ($P_{\mathrm{sticking}}$) can be defined as $P_{\mathrm{sticking}} = n_{\mathrm{dissociation}}/n_{\mathrm{all}}$, where $n_{\mathrm{dissociation}}$ is the number of trajectories that ended with dissociation, and $n_{\mathrm{all}}$ is the number of all the trajectories. 

\section*{Data Availability}
The H$_2$ on Cu(111) dataset and relevant simulation workflow scripts have previously been published and are publicly available.\cite{stark_machine_2023, stark_benchmarking_2024} The Ti-Al-V dataset, alongside the baseline Ti-Al-V ACE model we compare against, is made available in the dedicated repository: \url{https://zenodo.org/records/14244105}.

\section*{Code Availability}
The MACE-freeze patch to the MACE software suite is available under URL https://github.com/7radians/
mace-freeze/tree/mace-freeze.

\bibliographystyle{naturemag}
\bibliography{references}

\begin{acknowledgments}
The authors acknowledge financial support from the UKRI Future Leaders Fellowship program (MR/S016023/1, MR/X023109/1), a UKRI frontier research grant (ERC StG, EP/X014088/1), the CASTEP-USER grant funded by UK Research and Innovation (EP/W030438/1) and the EPSRC-funded Centre for Doctoral Training in Modelling of Heterogeneous Systems (HetSys CDT, EP/S022848/1).
High-performance computing resources were provided via the Scientific Computing Research Technology Platform of the University of Warwick and the EPSRC-funded HPC Midlands+ computing centre for access to Sulis (EP/P020232/1).
\end{acknowledgments}

\section*{Competing Interests}
R.J.M. is an associate editor for npj Computational Materials. All other authors declare no competing interests.

\end{document}



\title{Supplementary Information to \\``Fine-tuning foundation models of materials interatomic potentials with frozen transfer learning''}


\author{Mariia Radova}
\affiliation{Department of Chemistry, University of Warwick, Gibbet Hill Road, Coventry CV4 7AL, United Kingdom}
\affiliation{Department of Physics, University of Warwick, Gibbet Hill Road, Coventry CV4 7AL, United Kingdom}

\author{Wojciech G. Stark}%
\affiliation{Department of Chemistry, University of Warwick, Gibbet Hill Road, Coventry CV4 7AL, United Kingdom}

\author{Connor Allen}%
\affiliation{Department of Physics, University of Warwick, Gibbet Hill Road, Coventry CV4 7AL, United Kingdom}
\affiliation{Warwick Centre for Predictive Modelling, School of Engineering, University of Warwick, Gibbet Hill Road, Coventry CV4 7AL, United Kingdom}

\author{Reinhard J. Maurer}
\email{r.maurer@warwick.ac.uk}
\affiliation{Department of Chemistry, University of Warwick, Gibbet Hill Road, Coventry CV4 7AL, United Kingdom}
\affiliation{Department of Physics, University of Warwick, Gibbet Hill Road, Coventry CV4 7AL, United Kingdom}
\author{Albert P. Bart\'ok}
\email{Albert.Bartok-Partay@warwick.ac.uk}
\affiliation{Department of Physics, University of Warwick, Gibbet Hill Road, Coventry CV4 7AL, United Kingdom}
\affiliation{Warwick Centre for Predictive Modelling, School of Engineering, University of Warwick, Gibbet Hill Road, Coventry CV4 7AL, United Kingdom}

\maketitle

\section{Fine tuning of foundation models with delta learning\label{sec:1}}

Delta learning has been previously used in the context of machine learning interatomic potentials \cite{Westermayr2021} with a SchNet \cite{SchNet} model. This technique allows to learn the difference between a baseline (in this case, MACE-MP) method and a high accuracy method (DFT). The energies and forces of the relevant structures are replaced with the corresponding differences between their values predicted by the foundation models and the reference values, later referred to as ``deltas''. The delta models are small models trained on the deltas and their predictions are then added to the predictions of the baseline models:
\begin{equation}
\epsilon_{\textrm{pred}} = \epsilon_{\Delta} + \epsilon_{\textrm{MP}}
\end{equation}
\begin{equation}
\epsilon_{\Delta} = \epsilon_{\textrm{MP}} - \epsilon_{\textrm{DFT}}
\end{equation}

Delta learning is meant to be a cheap and efficient way of fine-tuning a machine learning model, with the assumption that the deltas can be fit, and the fitting function is simple enough so that smaller models with less number of parameters can learn the deltas.

\section{Supplementary figures}

\begin{figure}
    \centering
    \includegraphics[width=1\linewidth]{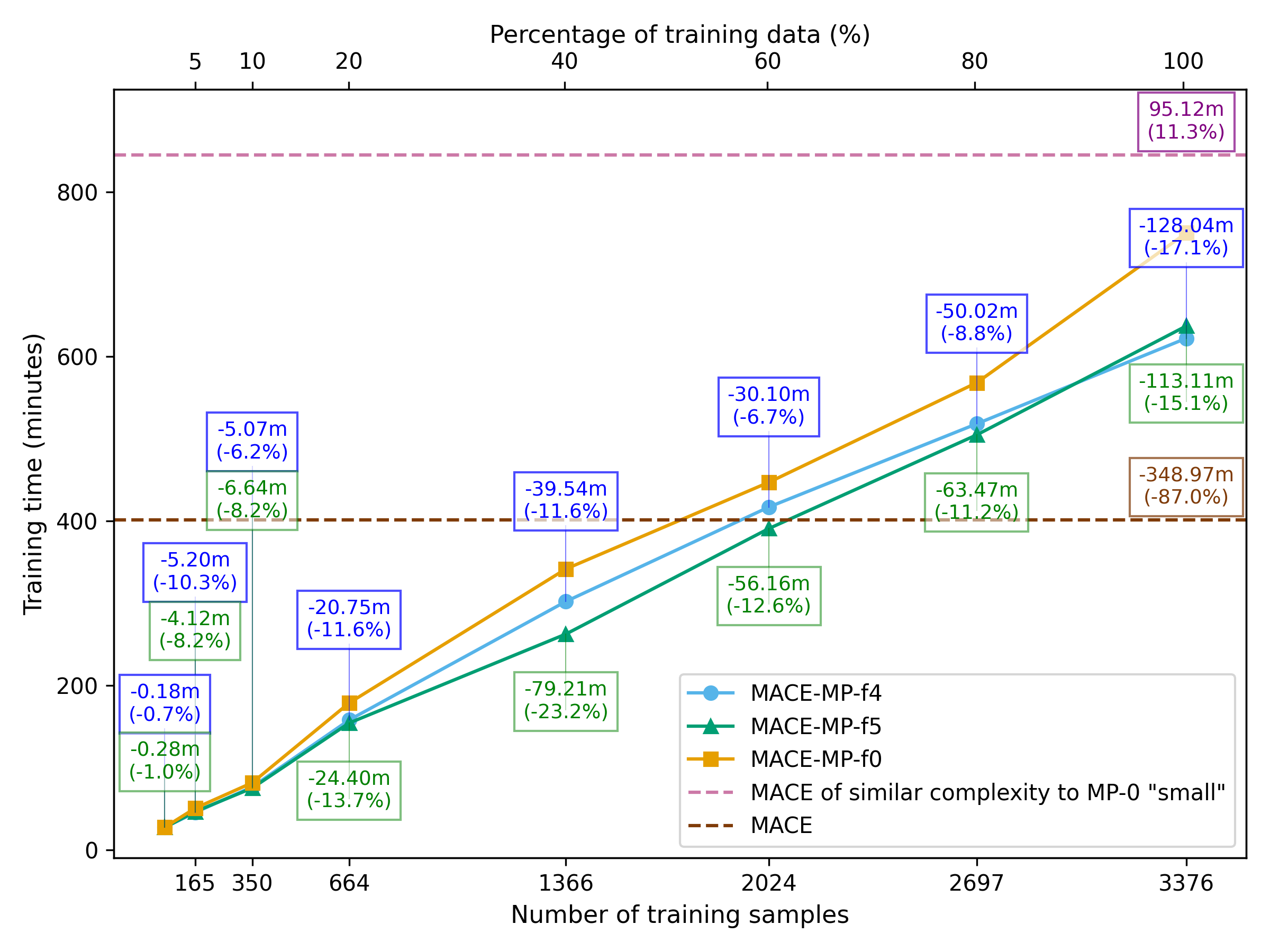}
    \caption{\textbf{Effective training time of Cu-H models.} Training times are shown for the MACE-MP-f4 (blue circles), MACE-MP-f5 (green triangles) and MACE-MP-f0 (yellow squares) transfer models, and MACE optimised (brown dashed line) and MACE of the similar size and complexity to MACE-MP-0 ``small'' (purple dashed line). The speedup is calculated relative to the MACE-MP-f0 models. This plot presents the effective training time for the Cu-H system, which is influenced by two primary factors: the acceleration of epoch completion due to parameter freezing and the time required to obtain the "best model" saved by MACE, given that all models are trained for the same number of epochs. Consequently, fluctuations in percentage values are expected. However, these variations should provide a representative estimate of the computational time savings achievable by the user.}
    \label{fig:speedup_hcu}
\end{figure}

\begin{figure}
    \centering
    \includegraphics[width=1\linewidth]{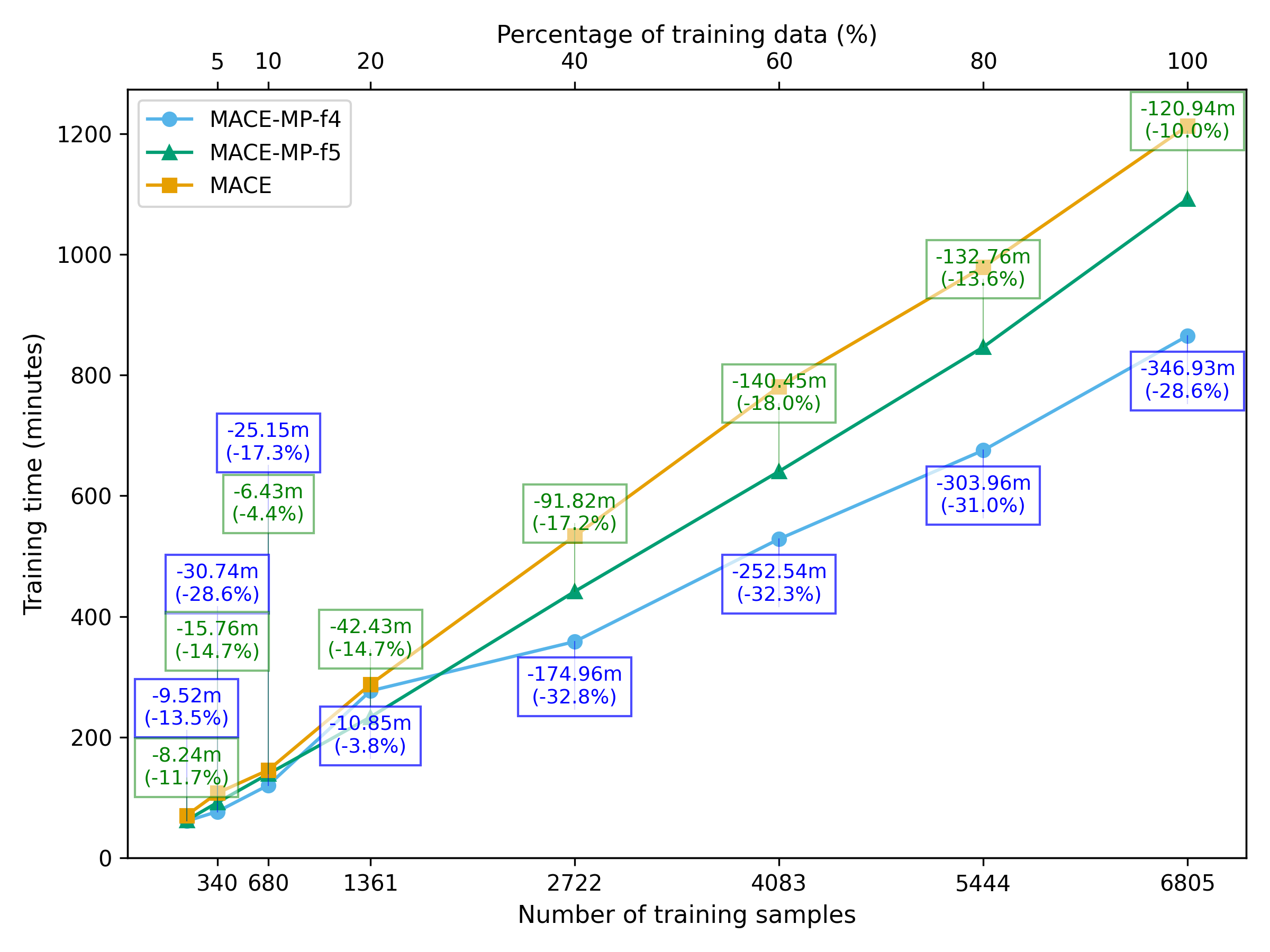}
    \caption{\textbf{Effective training time of Ti-Al-V models.} Training times are shown for the MACE-MP-f4 (blue circles), MACE-MP-f5 (green triangles) and MACE from-scratch (yellow squares). Here, the speedup is measured relative to the from-scratch MACE model of similar size to MACE-MP-0 ``small''. Notably, for the Ti-Al-V system, despite backpropagation being faster in the MACE-MP-f5 setting, convergence to the "best model" occurred more quickly with the MACE-MP-f4 setting. This behavior contrasts with the Cu-H system, where such a trend was not observed.
}
    \label{fig:speedup_ti}
\end{figure}

\begin{figure}
    \centering
    \includegraphics[width=1\linewidth]{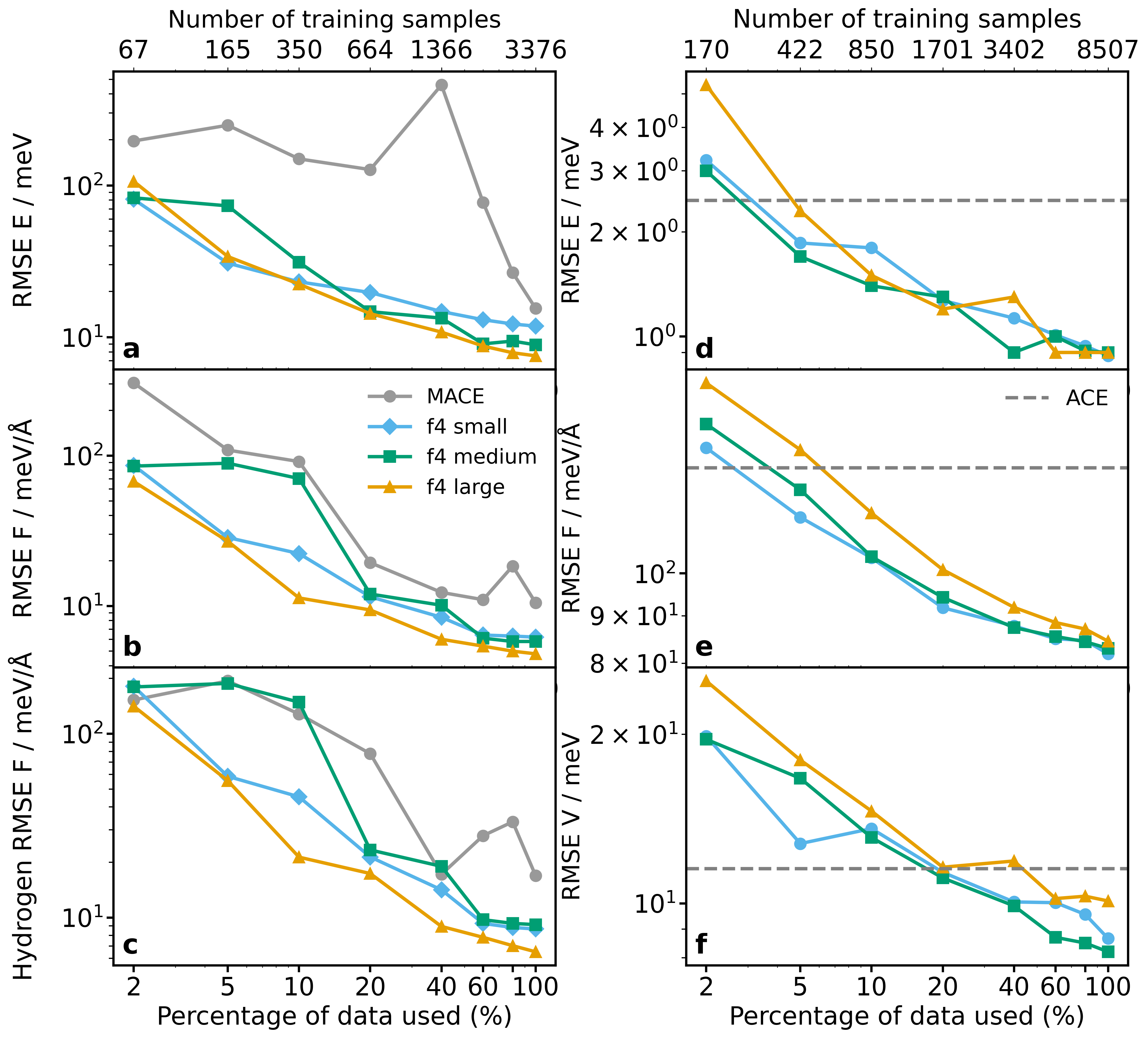}
    \caption{\textbf{Transfer learning curves for Hydrogen on Copper surface and the Ti-Al-V systems.}
     Panels (a), (b), (c) refer to the Cu-H$_2$ system, and panels (d), (e), (f) refer to the Ti-Al-V system. These transfer models were trained on MACE-MP ``small'' (blue diamonds), ``medium'' (green squares), and ``large'' (yellow triangles) foundation models. The gray circles mark the learning curve of the from-scratch MACE model. The gray dashed line marks the errors of the custom ACE model. The points correspond to the percentages of the full dataset, namely 2, 5, 10, 20, 40, 60, 80 and 100 \%. }
    \label{fig:lc_sizes}
\end{figure}

\begin{figure}
    \centering
    \includegraphics[width=3.3in]{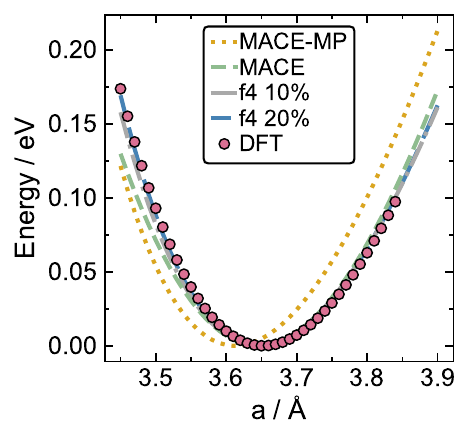}
    \caption{\textbf{Relative potential energies of the Cu atoms for a set of lattice constants}. Energies were calculated using DFT (red circles) and all the MLIPs included in our study, namely, MACE (green, shortly-dashed line), MACE-MP (yellow, dotted line), MACE-MP-f4 10\% (f4 10\%) (gray, ``dash-dot'' line), and MACE-MP-f4 20\% (f4 20\%) (blue, dashed line).}
    \label{fig:lattice}
\end{figure}

\begin{figure}
    \centering
    \includegraphics[width=3.3in]{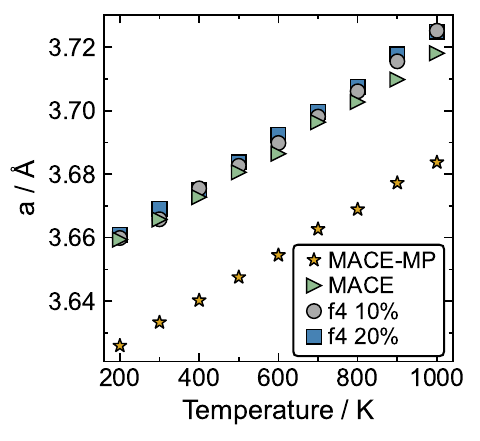}
    \caption{\textbf{Relation between the lattice constant of the Cu slab and temperature.}. The lattice constants were extracted from NPT simulations at 9 temperatures ranging from 200 and 1000~K, using MACE-MP (yellow stars), from-scratch MACE (green triangles), MACE-MP-f4 10\% (grey circles), and MACE-MP-f4 20\% (blue squares) models.}
    \label{fig:lattice_temp}
\end{figure}

\begin{figure*}
    \centering
    \includegraphics[width=1\linewidth]{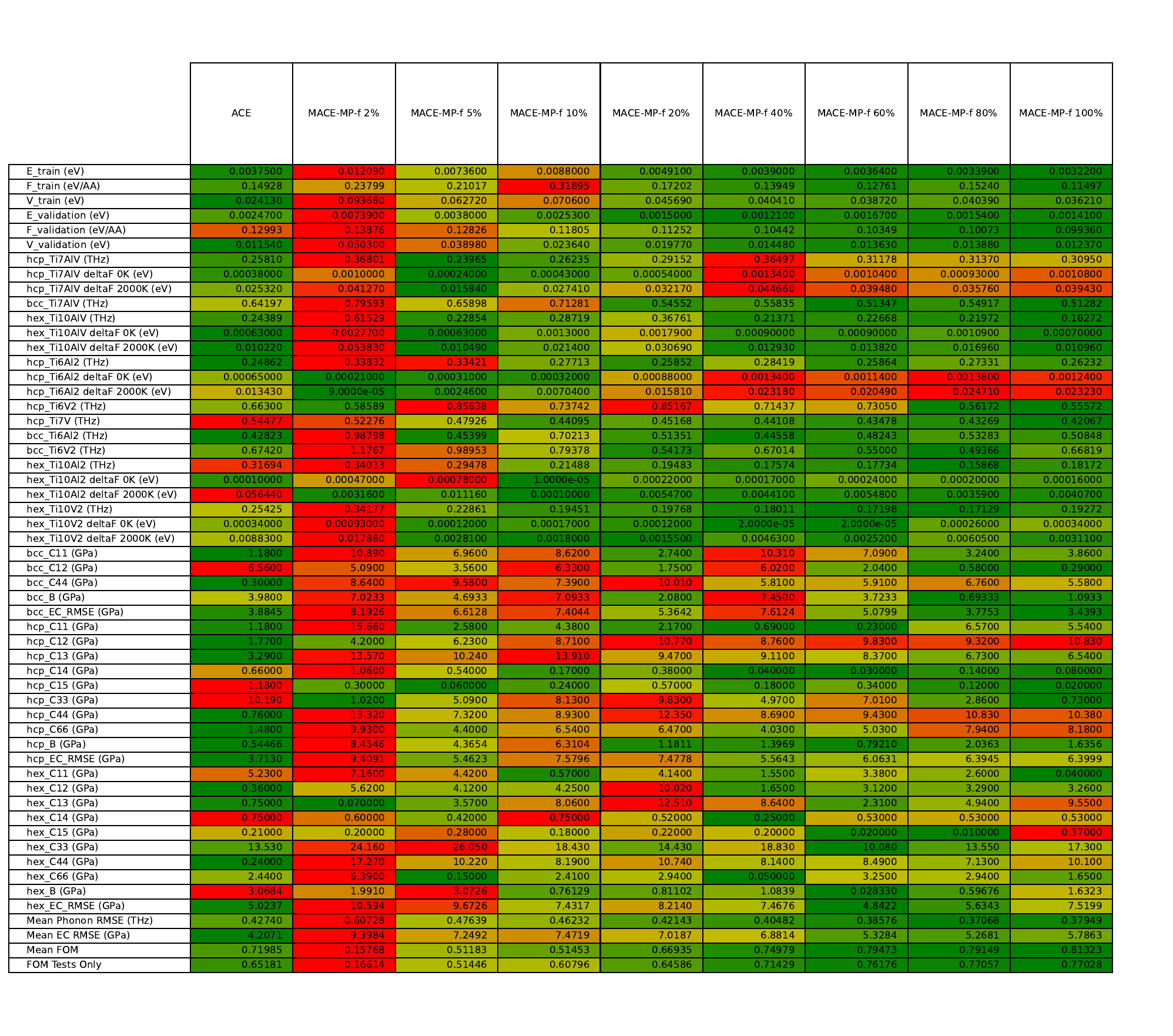}
    \caption{\textbf{Freeze=5 Ti-Al-V models evaluated on the Ti-Al-V benchmarks}. The results are summarised in the figure of merit (FOM), which measure the performance of models on all benchmarks, relative to their alternatives. The models are trained on the different \% of data, as outlined in the column titles. Details on the generation of this dataset can be found in  Ref.~\cite{allen2025multiphasedatasettiti6al4v}.}
    \label{fig:fom5_table}
\end{figure*}

\begin{figure*}
    \centering
    \includegraphics[width=1\linewidth]{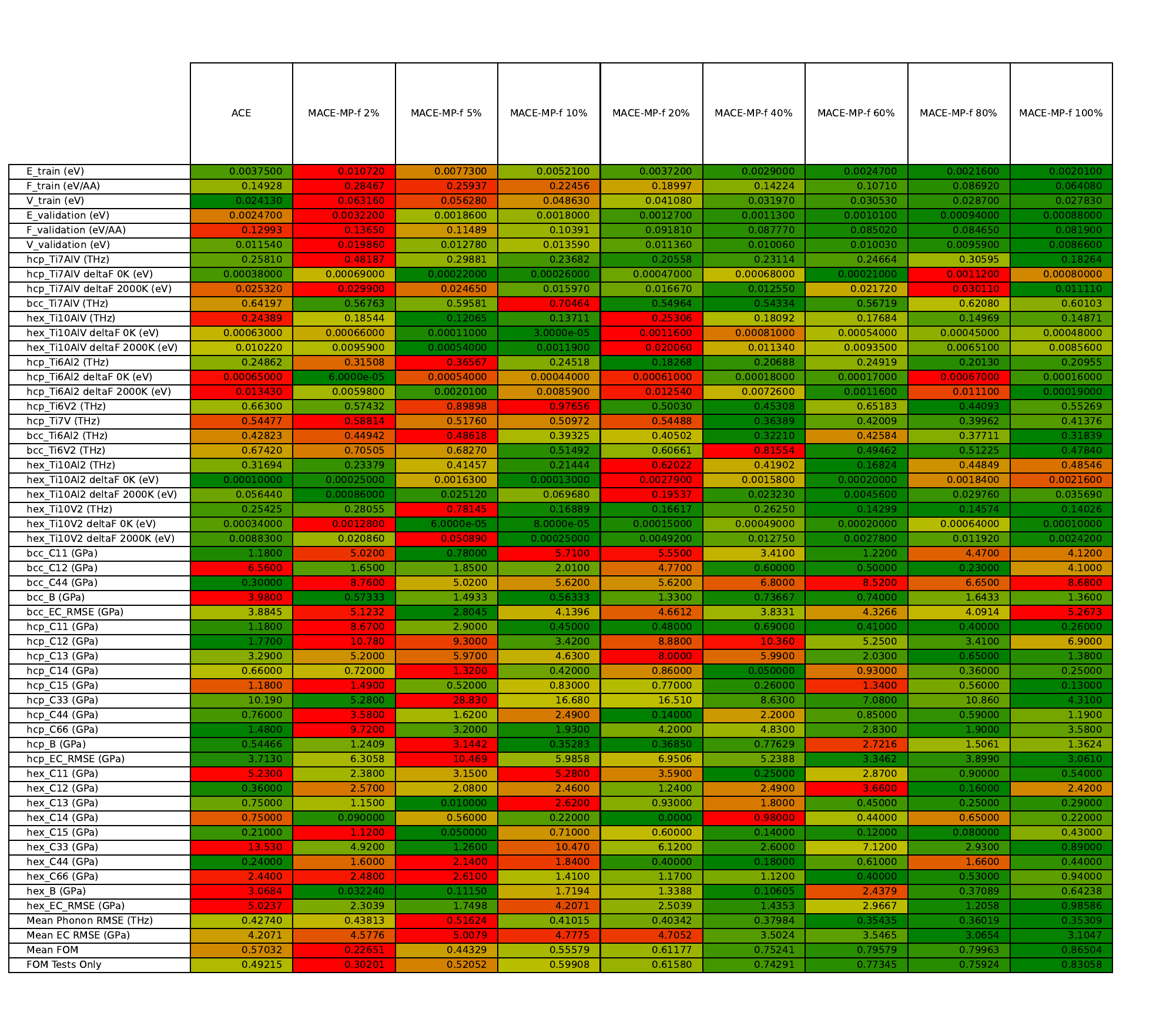}
    \caption{\textbf{Freeze=4 Ti-Al-V models evaluated on the Ti-Al-V benchmarks}. The results are summarised in the figure of merit (FOM), which measure the performance of models on all benchmarks, relative to their alternatives. The models are trained on the different \% of data, as outlined in the column titles.  Details on the generation of this dataset can be found in  Ref.~\cite{allen2025multiphasedatasettiti6al4v}.}
    \label{fig:fom4_table}
\end{figure*}

\begin{figure*}
    \centering
    \includegraphics[width=1\linewidth]{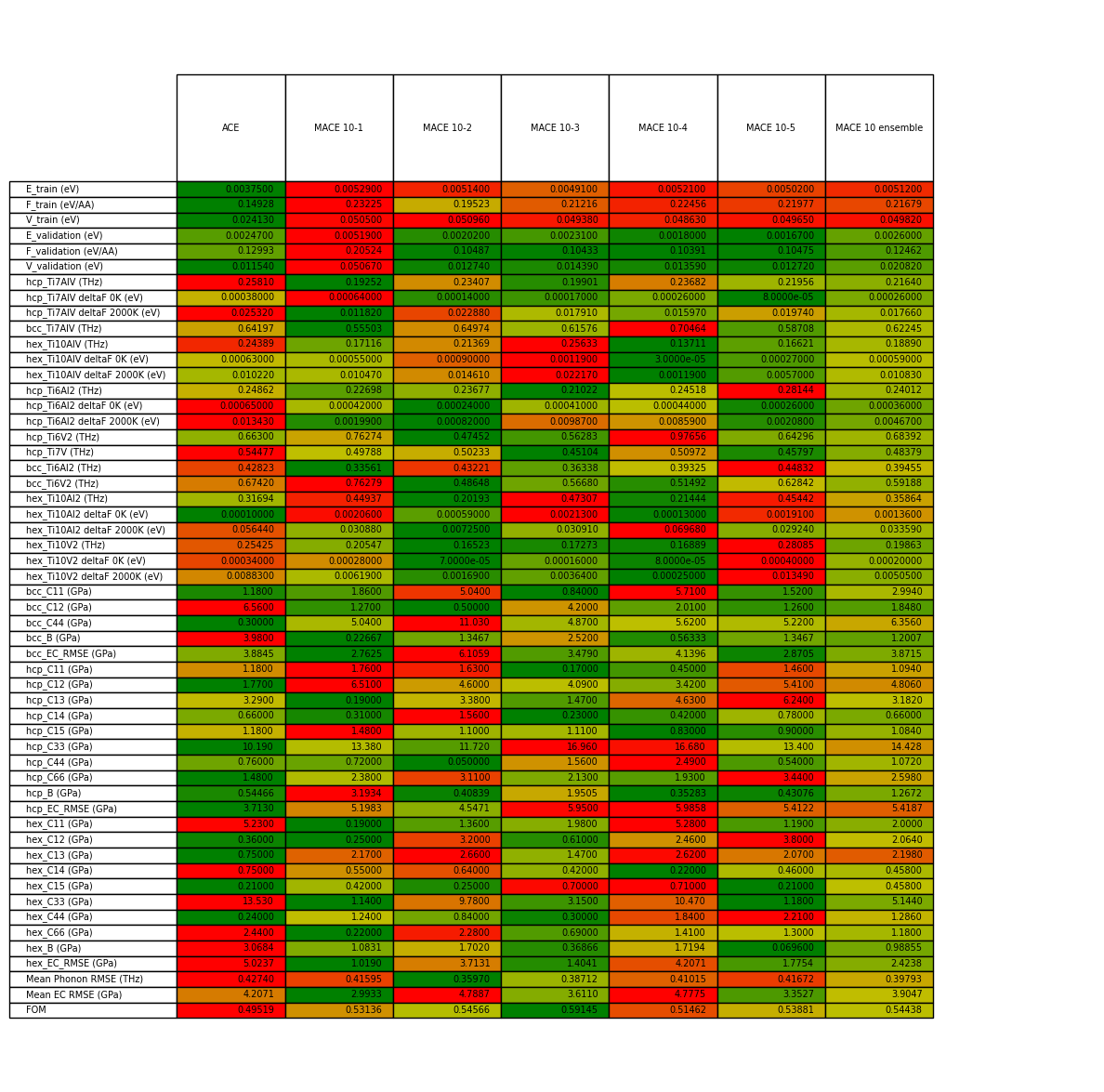}
    \caption{\textbf{Freeze=4 10\% ensemble of models compared with ACE-C, using an ensemble average}. This figure  quantified uncertainty of the models by training on the randomly uniformly sampled 10\% subsets of the full training set.  Details on the generation of this dataset can be found in Ref.~\cite{allen2025multiphasedatasettiti6al4v}.}
    \label{fig:ensemble-10}
\end{figure*}

\begin{figure*}
    \centering
    \includegraphics[width=1\linewidth]{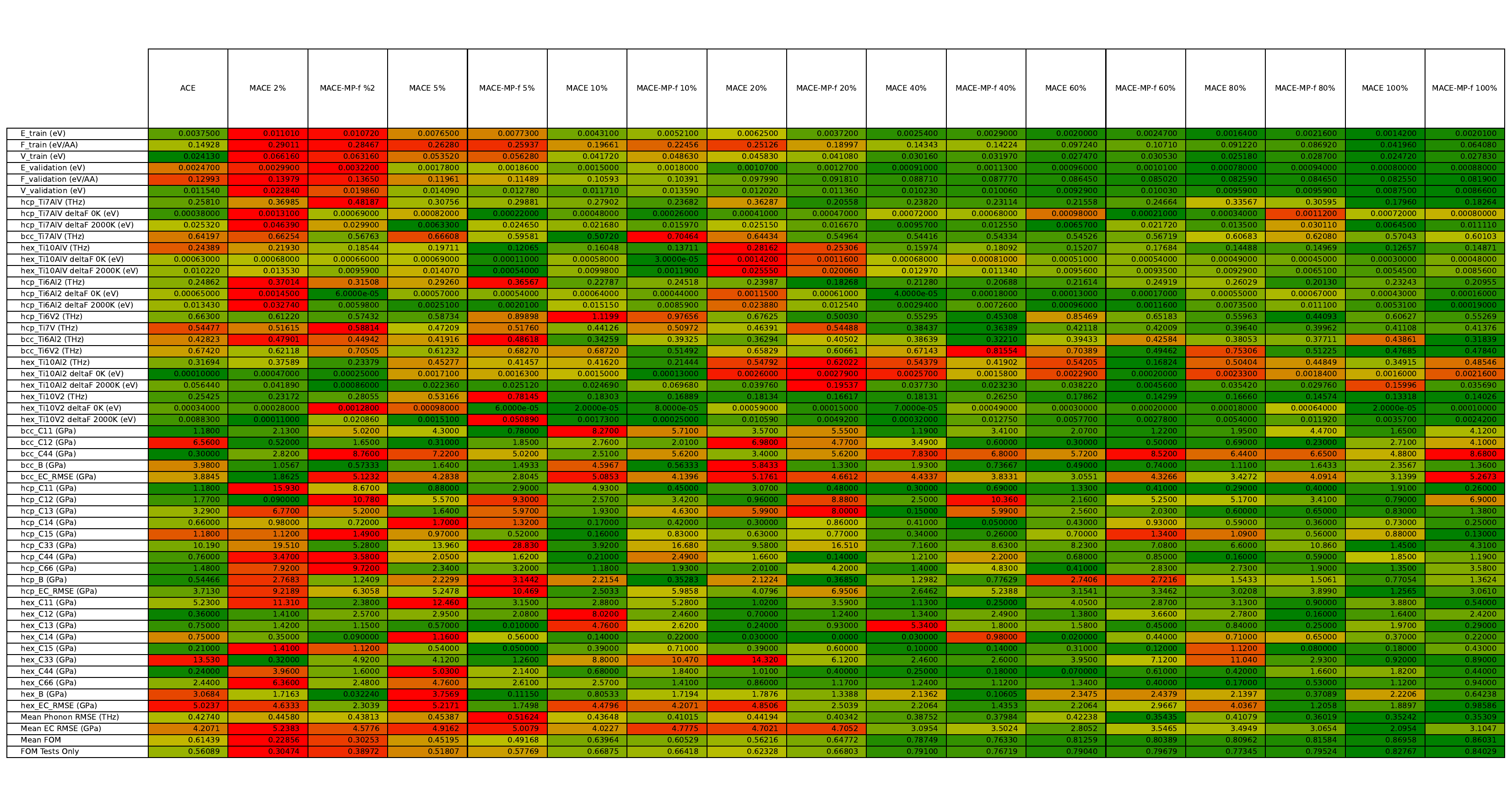}
    \caption{\textbf{Freeze=4 comparred with MACE from-scratch Ti-Al-V models evaluated on the Ti-Al-V benchmarks}. The results are summarised in the figure of merit (FOM), which measure the performance of models on all benchmarks, relative to their alternatives. The models are trained on the different \% of data, as outlined in the column titles.  Details on the generation of this dataset can be found in  Ref.~\cite{allen2025multiphasedatasettiti6al4v}.}
    \label{fig:freeze4-vs-from-scratch-table}
\end{figure*}

\begin{figure}
    \centering
    \includegraphics[width=3.3in]{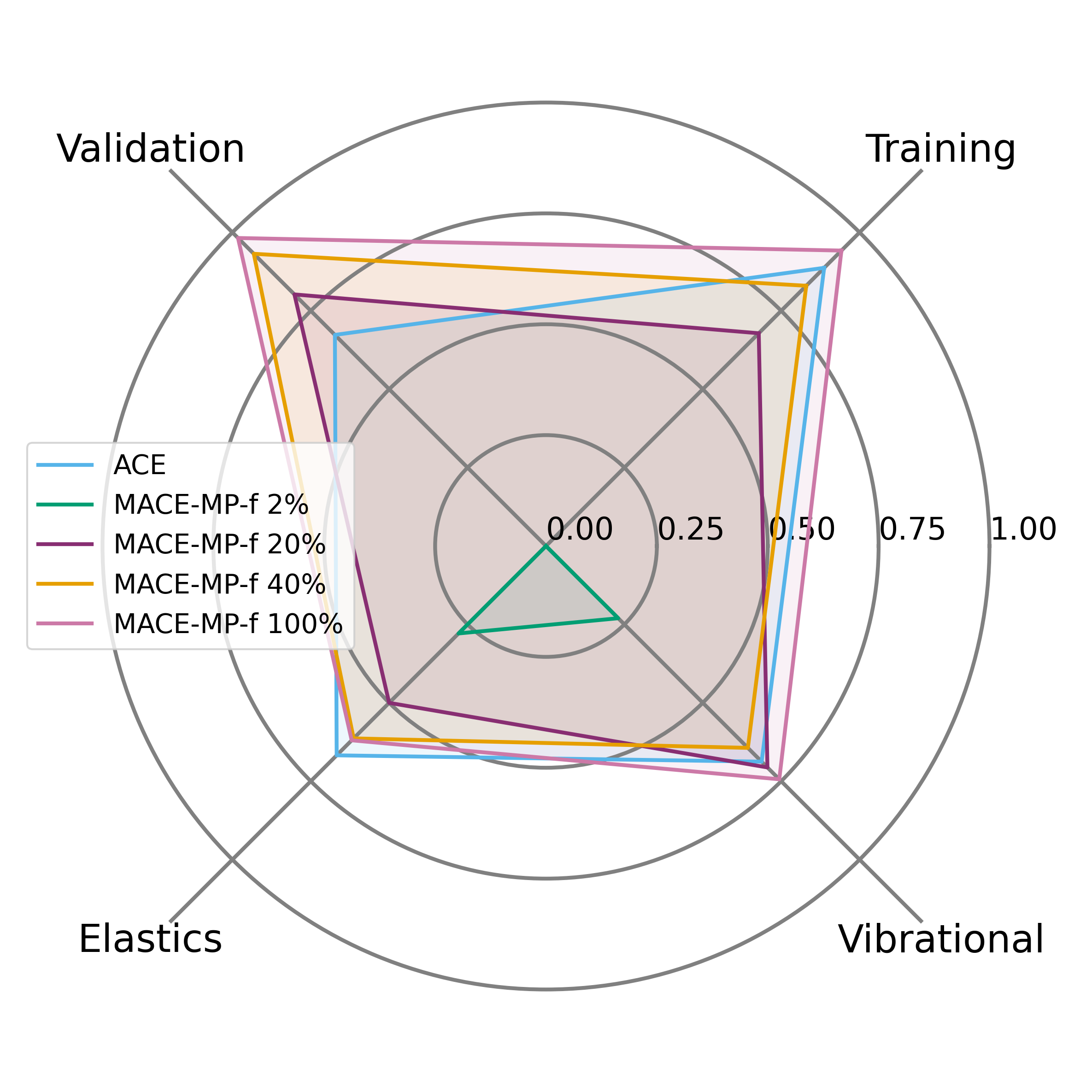}
    \caption{\textbf{Figures of merit (FOM) for freeze=5 models relative to the custom ACE}. This figure summarises the comparison of predictive performance across a list of benchmarks in Fig.\ref{fig:fom5_table}. }
    \label{fig:foms5}
\end{figure}

\begin{figure}
    \centering
    \includegraphics[width=6.6in]{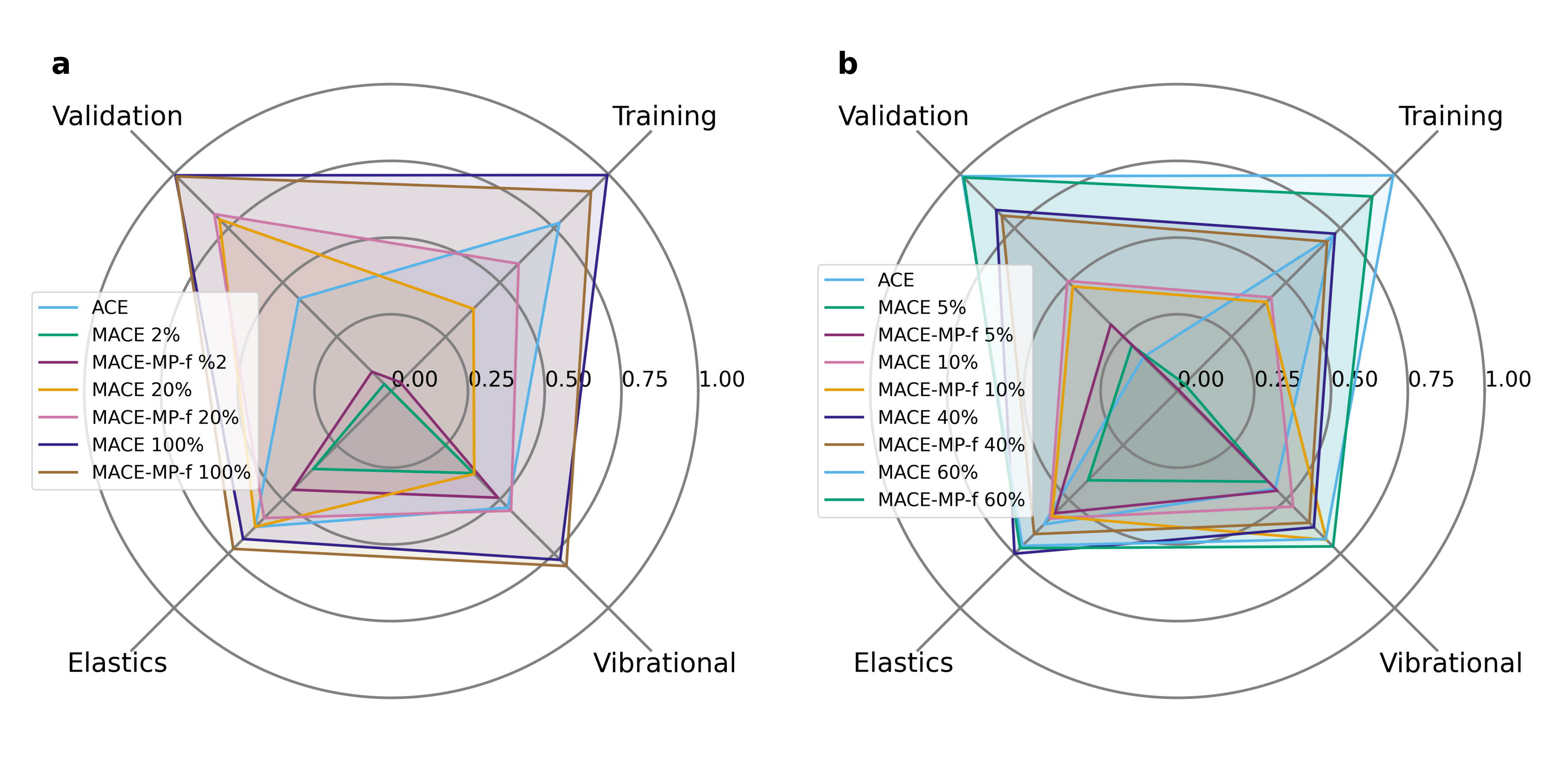}
    \caption{\textbf{Figures of merit (FOM) for freeze=4 models against the from-scratch MACE models.}: 
    Panel (a) shows the training subsets of 2, 20 and 100\%. Panel (b) shows training subsets of 5, 10, 40 and 60\%. These figures summarise the comparison of predictive performance across a list of benchmarks in Fig.\ref{fig:freeze4-vs-from-scratch-table}. }
    \label{fig:from-scratch-radar}
\end{figure}

\begin{figure}
    \centering
    \includegraphics[width=3.3in]{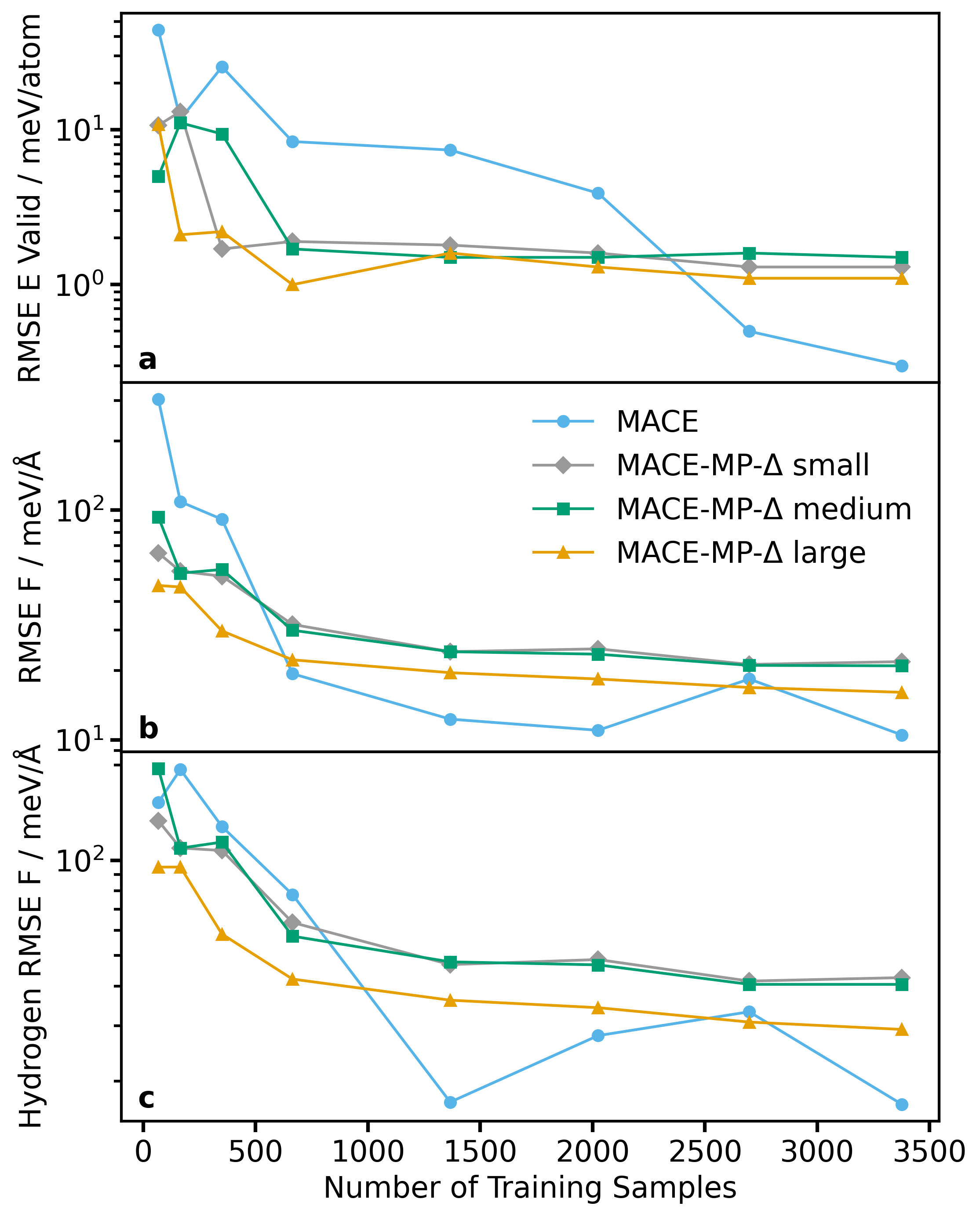}
    \caption{\textbf{Delta learning curves for Hydrogen on Copper surface}. These delta models were trained based on MACE-MP ``small'' (grey diamonds), ``medium'' (green squares), and ``large'' (yellow triangles) foundation models. The ``MACE'' label (blue circles) marks the learning curve of the from-scratch MACE model. The points correspond to the percentages of the full dataset, namely 2, 5, 10, 20, 40, 60, 80 and 100 \%.  }
    \label{fig:lc_delta}
\end{figure}

\begin{figure}
    \centering
    \includegraphics[width=3.9in]{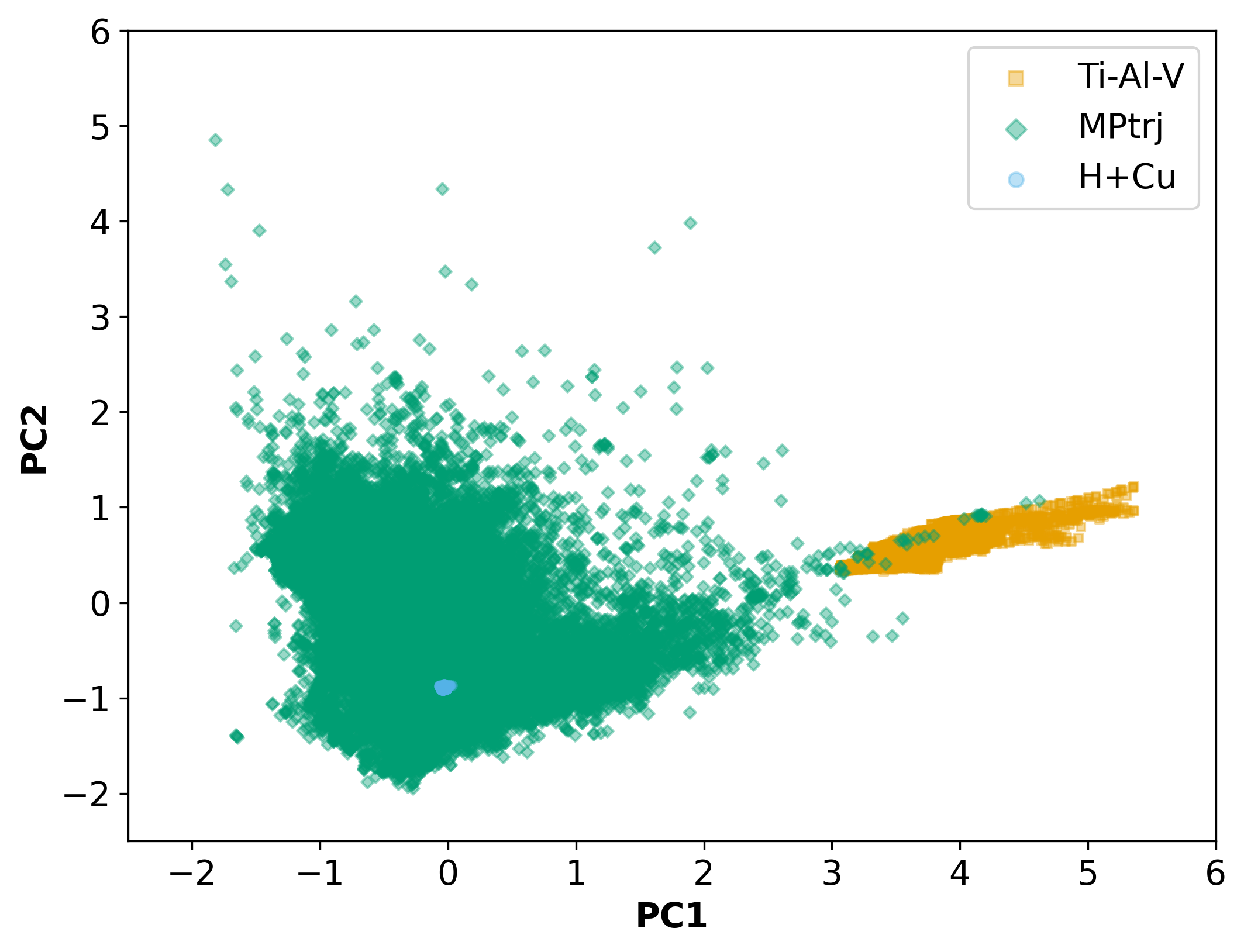}
    \caption{\textbf{Principal Component Analysis (PCA) of the MACE-MP-0 ``small'' features (invariant descriptors).} Points correspond to configurations from MPtrj (green diamonds), Cu-H (blue circles), Ti-Al-V (yellow squares).}
    \label{fig:pca}
\end{figure}

\clearpage
\bibliographystyle{naturemag}
\bibliography{references}
